\documentclass[sigconf]{acmart} 
\usepackage{xspace}
\usepackage{makecell}
\usepackage{url}
\usepackage{soul}
\usepackage[caption=false,font=footnotesize]{subfig}
\DeclareUnicodeCharacter{2212}{-}

\definecolor{DarkGreen}{HTML}{5DAC81}

\newcommand{\system}{SepsisLab\xspace}
\AtBeginDocument{%
  \providecommand\BibTeX{{%
    \normalfont B\kern-0.5em{\scshape i\kern-0.25em b}\kern-0.8em\TeX}}}

\copyrightyear{2024}
\acmYear{2024}
\setcopyright{rightsretained}
\acmConference[CHI '24]{Proceedings of the CHI Conference on Human Factors in Computing Systems}{May 11--16, 2024}{Honolulu, HI, USA}
\acmBooktitle{Proceedings of the CHI Conference on Human Factors in Computing Systems (CHI '24), May 11--16, 2024, Honolulu, HI, USA}
\acmDOI{10.1145/3613904.3642343}
\acmISBN{979-8-4007-0330-0/24/05}

\begin{document}


\title{Rethinking Human-AI Collaboration in Complex Medical Decision Making: A Case Study in Sepsis Diagnosis}

\author{Shao Zhang}
\email{zhang.shao1@northeastern.edu}
\orcid{0000-0002-0111-0776}
\affiliation{%
  \institution{Northeastern University}
  \city{Boston}
  \state{Massachusetts}
  \country{United States}
}

\author{Jianing Yu}
\email{jiani.yu@northeastern.edu}
\orcid{0009-0006-6197-4812}
\affiliation{%
  \institution{Northeastern University}
  \city{Boston}
  \state{Massachusetts}
  \country{United States}
}

\author{Xuhai Xu}
\email{xoxu@mit.edu}
\orcid{0000-0001-5930-3899}
\affiliation{%
  \institution{Massachusetts Institute of Technology}
  \city{Cambridge}
  \state{Massachusetts}
  \country{United States}
}

\author{Changchang Yin}
\email{yin.731@osu.edu}
\orcid{0000-0002-6540-6365}
\affiliation{%
  \institution{The Ohio State University}
  \city{Columbus}
  \state{Ohio}
  \country{United States}
}

\author{Yuxuan Lu}
\email{lu.yuxuan@northeastern.edu}
\orcid{0000-0002-8520-0540}
\affiliation{%
  \institution{Northeastern University}
  \city{Boston}
  \state{Massachusetts}
  \country{United States}
}

\author{Bingsheng Yao}
\email{yaob@rpi.edu}
\orcid{0009-0004-8329-4610}
\affiliation{%
  \institution{Rensselaer Polytechnic Institute}
  \city{Troy}
  \state{New York}
  \country{United States}
}

\author{Melanie Tory}
\email{m.tory@northeastern.edu}
\orcid{0000-0002-6806-9253}
\affiliation{%
  \institution{Northeastern University}
  \city{Portland}
  \state{Maine}
  \country{United States}
}

\author{Lace M. Padilla}
\email{l.padilla@northeastern.edu}
\orcid{0000-0001-9251-5279}
\affiliation{%
  \institution{Northeastern University}
  \city{Boston}
  \state{Massachusetts}
  \country{United States}
}

\author{Jeffrey Caterino}
\email{Jeffrey.Caterino@osumc.edu}
\orcid{}
\affiliation{%
  \institution{The Ohio State University Wexner Medical Center}
  \city{Columbus}
  \state{Ohio}
  \country{United States}
}

\author{Ping Zhang}
\authornote{Correspondence to Ping Zhang and Dakuo Wang}
\email{zhang.10631@osu.edu}
\orcid{0000-0002-4601-0779}
\affiliation{%
  \institution{The Ohio State University}
  \city{Columbus}
  \state{Ohio}
  \country{United States}
}

\author{Dakuo Wang}
\authornotemark[1]
\email{d.wang@northeastern.edu}
\orcid{0000-0001-9371-9441}
\affiliation{%
  \institution{Northeastern University}
  \city{Boston}
  \state{Massachusetts}
  \country{United States}
}

\renewcommand{\shortauthors}{Zhang et al.}

\begin{abstract}
Today's AI systems for medical decision support often succeed on benchmark datasets in research papers but fail in real-world deployment. This work focuses on the decision making of sepsis, an acute life-threatening systematic infection that requires an early diagnosis with high uncertainty from the clinician. Our aim is to explore the design requirements for AI systems that can support clinical experts in making better decisions for the early diagnosis of sepsis.
The study begins with a formative study investigating why clinical experts abandon an existing AI-powered Sepsis predictive module in their electrical health record (EHR) system. 
We argue that a human-centered AI system needs to support human experts in the intermediate stages of a medical decision-making process (e.g., generating hypotheses or gathering data), instead of focusing only on the final decision.
Therefore, we build \system based on a state-of-the-art AI algorithm and extend it to predict the future projection of sepsis development, visualize the prediction uncertainty, and propose actionable suggestions (i.e., which additional laboratory tests can be collected) to reduce such uncertainty.
Through heuristic evaluation with six clinicians using our prototype system, we demonstrate that \system enables a promising human-AI collaboration paradigm for the future of AI-assisted sepsis diagnosis and other high-stakes medical decision making.

\end{abstract}



\begin{teaserfigure}
    \centering
    \vspace{-0.3cm}
    \includegraphics[width=0.9\linewidth]{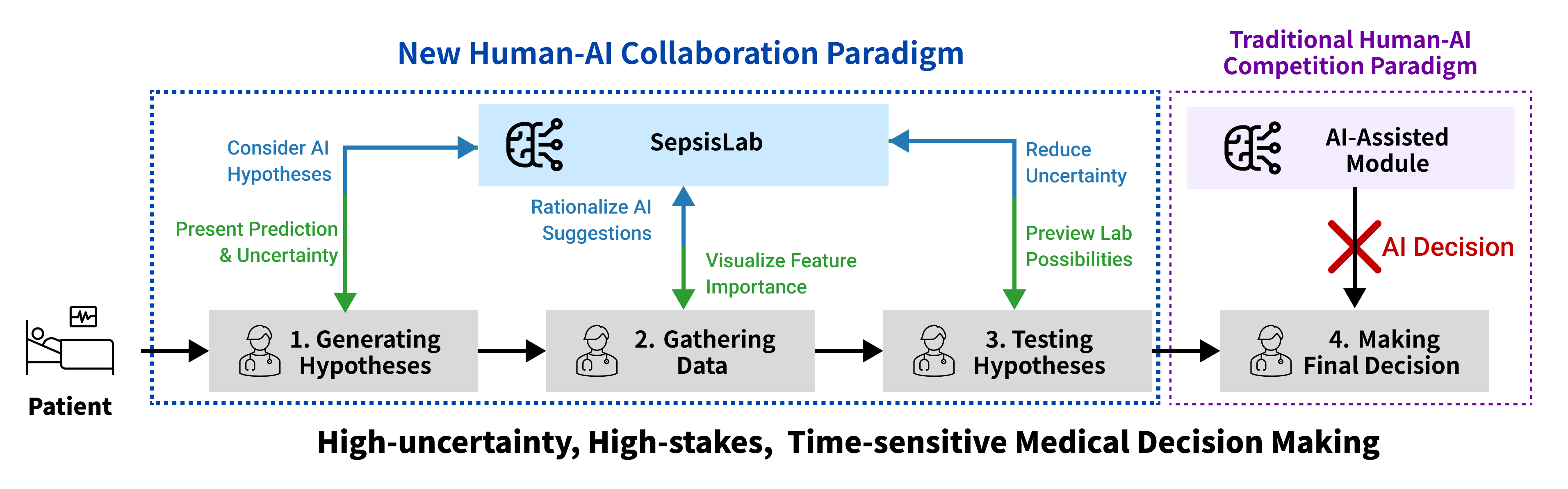}
    \vspace{-0.2cm}
    \caption{\textbf{Existing EPIC Sepsis Module and Our Proposed Sepsis Decision-Support Module in Medical Decision Making Workflow}. Our work focuses on sepsis diagnosis, a high-uncertainty, high-stakes, time-sensitive medical decision-making process. Physicians usually take four steps: (1) generating hypotheses, (2) gathering data, (3) testing hypotheses, and (4) making final decisions. Our study results point out that existing sepsis module is not helpful, forming a human-AI competition paradigm. We propose a new module to establish a human-AI collaboration paradigm.}
    \label{fig:human-AI-collab-paradigm}
\end{teaserfigure}

\maketitle


\section{Introduction}

There is a growing interest from both academia and industry in the development of artificial intelligence (AI) to support medical decision making~\cite{yu2018artificial,cai2019hello,gutierrez2020artificial,lentzen2022critical,lai2021towards,xu2019leveraging,lu2022contextual,9765710}.
Although the target scenarios may vary, from diagnostic decision making with medical imaging ~\cite{ting2018ai} to outpatient symptom triage~\cite{wang2021brilliant}, the ultimate goal remains the same: to reduce the burden of human medical experts while improving the quality of the final decision.
Along this direction, many novel deep learning-based AI algorithms have been proposed, most of which yield promising predictive performance in their corresponding benchmark datasets~\cite{cheng2016risk,baytas2017patient,zhu2016measuring}, and some of them even outperform human experts in head-to-head competitions within controlled experimental settings~\cite{zhang2021interpretable}.

However, the deployments of these AIs face more resistance in reality than their promising accuracy scores reported in research papers \cite{wong2021external,topol2019high,rajkomar2019machine,xu2022globem,xu2023globem}.
Luckily, more and more researchers have recently noticed the growing number of failure cases where AI-assisted medical decision-making systems are being abandoned by their target users. Recently, researchers have conducted various empirical studies to explore the cause of unsuccessful human-AI collaborative decision making \cite{shah2019ai,yang2016investigating,yang2019unremarkable,romero2020lesson,ghassemi2021false,amann2022explain,chaddad2023survey, otaki2022clinical,10.1145/3313831.3376590,10.1145/3313831.3376624}.
For example, human experts are the only ones responsible for an inaccurate diagnosis, while AI is not, so the clinicians trust their own judgment more than AI prediction \cite{xiong2022challenges,wang2021brilliant}.
Based on these findings, they have proposed various suggestions for user interface and user experience design (for example, to improve physician adoption of AI with new eXplainable AI (XAI) features~\cite{ghassemi2021false,amann2022explain,10.1145/3313831.3376590,10.1145/3313831.3376624,chaddad2023survey,otaki2022clinical,zhang2020effect}).

In this work, we join the research effort to design AI to support medical decision making while focusing on the scenario of \textbf{sepsis diagnosis}. Sepsis is a common (~48.9 million patients per year worldwide) yet life-threatening organ dysfunction triggered by a dysregulated response to infection~\cite{world2020global}. 
The development of sepsis is very fast --- without a timely diagnosis or proper treatment, a patient might die within a few hours from the initial onset of symptoms~\cite{zambon2008implementation}.
Compared to other medical decision-making scenarios (e.g., abnormal cell detection in medical imaging) where clinicians make a decision for that particular moment and with all the information they have at hand (e.g., cancer cells present or absent in the image), sepsis diagnosis is particularly challenging because:
1) clinicians need to decide not only whether the patient has sepsis at that moment, but also how likely this patient may develop sepsis in the near future (e.g., in a few hours);
2) they often do not have enough information to support their decision making. 
For example, the golden standard test for sepsis is the ``blood culture'' test in the Sepsis-3 guidelines~\cite{singer2016third}. 
However, it takes at least 8 hours to obtain the result, which would most likely be too late for a patient with sepsis~\cite{dierig2018time,evans2021surviving}.
The early diagnosis of sepsis represents a common but under-explored decision-making scenario in the real world: it requires human experts with \textbf{specific domain knowledge} to cope with\textbf{ high-uncertainty }and to make a \textbf{high-stakes} and \textbf{time-sensitive} decision based on \textbf{insufficient information}.

These special characteristics of sepsis diagnosis pose novel challenges for an AI system designed to support such decision making.  
There are some early efforts of research work that aim to design AI-based solutions to support sepsis diagnosis~\cite{goh2021artificial,liu2019data, yuan2020development,yin2022deconfounding,sivaraman2023ignore,zhang2021interpretable,kim2019sepsis,cecconi2018sepsis,singer2016third,evans2021surviving,sendak2020human}.
One notable effort is that a number of hospitals recently started to adopt a sepsis prediction and alerting module,\textbf{ Epic Sepsis Module (ESM)}, in their existing Electrical Health Record (EHR) system~\cite{cull2023epic}.
The core of this sepsis module is a machine learning algorithm, which can take in a patient's EHR information and other biomarker data at that moment as input and predict a sepsis risk score as output~\cite{nguyen2014automated}.
When the predicted risk score is higher than a threshold, it suggests to human clinicians that a sepsis case presents, and human clinicians can agree, disagree, or dismiss such AI decision-making suggestions.
Among the \textbf{four stages of the medical decision-making workflow} (\textit{1. Generating hypotheses, 2. Gathering data, 3. Testing hypotheses, 4. Making decisions})~\cite{sox2013medical}, this AI system is designed to provide support for the \textit{final diagnosis} stage.

What do human experts think about such an AI-based sepsis diagnosis support system? 
Our work bridges this research gap with an interview study with six experienced clinicians who actively engage in sepsis diagnosis every day, and their hospital has recently adopted AI-based ESM technology.
The results reveal that human experts believe the current AI module to be useless or even an intimidating ``competitor'' for the targeted high uncertainty, high-stakes, and time-sensitive decision-making scenario of sepsis early diagnosis (the bottom of Figure~\ref{fig:human-AI-collab-paradigm}).
Based on the findings, we designed \textbf{\system} with the goal of supporting the earlier stages of a medical decision making workflow for sepsis diagnosis (\textit{hypotheses generation} and \textit{data gathering for hypotheses testing}) instead of making a blunt prediction for the \textit{final diagnosis decision}, as shown in the top of Figure~\ref{fig:human-AI-collab-paradigm}.
Our \system system can predict and visualize the likelihood and uncertainty range of whether a patient has sepsis at the moment, and whether they may develop sepsis in the near future.
In addition, \system can further suggest the most important but currently missing laboratory tests as an actionable suggestion for human clinicians, so that more data can be gathered to reduce uncertainty, leading to a more informed and higher quality final decision. 
A follow-up user evaluation study suggests that human experts appreciate our design, and they believe \system provides a better human-AI team experience compared to the existing human-AI competition paradigm.
We envision our findings within the sepsis diagnosis example can shed light on the design of more human-AI collaboration paradigms for other domain-specific, uncertain, high-stakes, and time-sensitive decision-making tasks.

This paper demonstrates the following contributions:
\begin{itemize}
    \item We conducted an empirical study to understand how human clinical experts interact with and perceive an existing AI-powered sepsis prediction module in their day-to-day work.
    \item We designed a new AI-assisted decision-making system, \system, which can predict and visualize the current and future likelihood and uncertainty of the onset of sepsis in a patient, and suggest actionable laboratory test recommendations to help human experts reduce the uncertainty of the final decision.
    \item We followed up with a user evaluation study, in which participants expressed a strong interest in adopting our system in their day-to-day work, and they believed our AI is no longer an ``intimidating competitor'' but more of a ``collaborator''.
\end{itemize}

\section{Background and Related Work}
\label{sec:related_work}

\subsection{Sepsis Diagnosis} 
\label{sub:related_work:sepsis}

Sepsis is a severe, life-threatening condition that affects approximately 48.9 million patients worldwide, resulting in around 11 million sepsis-related deaths \cite{cecconi2018sepsis,world2020global,rudd2020global}.
\citet{world2020global} highlighted the importance of early detection of sepsis symptoms and signs, along with the identification of biomarkers, for effective management.
In modern clinical practice, the Sepsis-3 guidelines \cite{singer2016third} serve as the gold standard for clinicians' diagnostic decisions.
In the early diagnosis of sepsis, clinicians rely on clinical evaluations, laboratory test results, and blood cultures \cite{evans2021surviving}.
The diagnosis process represents a common, complex, but under-explored decision-making scenario: It is high-stakes (life-threatening), time-sensitive (a patient's severe symptoms developed only in a few hours), and highly uncertain (need extensive lab test outcomes that they often don't have at the moment of decision-making).

We contextualize the 4-step process in \cite{sox2013medical} for the current complex sepsis diagnosis workflow adopted by clinicians:
(1) \textit{Generating Hypotheses}. Physicians evaluate sepsis-risk patients and form hypotheses by the information from EHR system and physical examination but under significant uncertainty.
(2) \textit{Gathering Data.} Physicians order lab tests based on the most promising hypotheses to gather more information.
(3) \textit{Testing Hypotheses}. Based on lab test outcomes, physicians refine or expand their hypotheses.
(4) \textit{Making Decisions.} Physicians diagnose based on the revised hypotheses.
As we show in Section \ref{sec:result}, our formative study results validate these steps.

With the rapid growth in volume and diversity of Electronic Health Records (EHRs), AI-driven algorithms have been studied for the sepsis onset risk prediction task. 
Screening tools have been used clinically to recognize sepsis, such as quick Sequential (Sepsis-Related) Organ Failure Assessment (qSOFA) \cite{singer2016third}, Modified Early Warning Score (MEWS) \cite{subbe2001validation}, National Early Warning Score (NEWS) \cite{smith2013ability}, and Systemic Inflammatory Response Syndrome (SIRS) \cite{bone1992definitions}. 
However, those tools were designed to screen existing symptoms, as opposed to explicitly predicting sepsis prior to its onset, and their efficacy in sepsis diagnosis is limited. 
For example, prior studies show that qSOFA had low sensitivities in identifying sepsis in both prehospital and emergency department (ED) settings \cite{dorsett2017qsofa,usman2019comparison}. 
In addition, deep-learning-based models are proposed to make sepsis onset predictions~\cite{rafiei2021ssp,lin2018early,zhang2017lstm,saqib2018early}. 
Recent studies have employed attention mechanisms to explain models' inner workings~\cite{zhang2021interpretable,kamal2020interpretable}.

However, despite the advantages of AI models' performance, these methods still often fail to garner clinician confidence, thereby hindering their practical implementation in real-world clinical settings~\cite{price2018big,duran2021afraid,10.1145/3491101.3503727,lee2021included}. 
For example, the Epic Sepsis Module (ESM) is the most widely used AI-based technique in current sepsis-related decision support practice~\cite{cull2023epic}.
The patient's data and lab test results (if any) are fed into the ESM module, generating a risk score. If the score is above a threshold, an alert will be sent to physicians and nurses, aiming to help with Step (4) in the current workflow.
Yet a large number of studies have shown that its effect is impacted by a large number of external factors and cannot stably improve patient treatment effects \cite{lyons2023factors,wong2021external}.
Existing HCI research on AI-CDSS in sepsis context primarily focuses on exploring treatment strategy choices during the treatment process \cite{sivaraman2023ignore} and investigating the transparency and explainability of AI algorithms \cite{sendak2020human}, but do not delve into the challenges of decision-making during the diagnostic phase.
Little is known about what human experts think about such an AI-based sepsis module in their diagnosis decision-making process.
Our work aims to address this gap by interviewing experienced clinicians.

\subsection{Challenges of AI-empowered Clinical Decision-Making Support}
\label{sub:related_work:challenge}

With AI's advancement, algorithms have been crafted to bolster clinical decision making, aiming to improve patient outcomes \cite{sivaraman2023ignore} and reduce clinician workload \cite{iftikhar2020artificial}.
Numerous studies indicate that AI-supported clinical decision support systems (AI-CDSS) can effectively assist doctors. AI's suggestions can prompt doctors to reflect deeper on a patient's condition and alert them to potential disease progression \cite{cai2019human,lee2021human,tanguay2022evaluating}.
For example, \citet{yang2019unremarkable} noted that clinical order recommendation systems have garnered positive feedback, with doctors asserting that such recommendations enhance their work efficiency.
\citet{caballero2017web} quantitatively demonstrated that incorporating AI could diminish the time doctors spend evaluating patients.

However, a significant hurdle for AI-supported clinical decision making is liability.
Given that clinicians bear the responsibility for medical decisions, they approach AI system predictions with utmost caution \cite{beede2020human,wang2021brilliant,lai2019human}.
The opaque nature of AI algorithms makes it challenging for doctors to fully embrace AI's direct diagnostic and treatment suggestions \cite{romero2020lesson,lu2021human}.
Recent research suggests that embedding AI into EHR systems might inadvertently increase physician workload \cite{juluru2021integrating,beede2020human}.
Furthermore, a disconnection exists between what clinicians expect from AI and what AI actually delivers \cite{cai2019hello,tanguay2022evaluating,lee2018understanding,10.1145/3582430}.
Some research also found that current AI-assisted decision-making does not align with clinicians decision-making in depressive disorder  \cite{10.1145/3411764.3445385} and type 2 diabetes \cite{10.1145/3544548.3581251} diagnosis process.
\citet{yang2016investigating} highlight that clinicians often adopt a `wait and see' approach, seeking evidence to validate their hypotheses before deciding.
In contrast, AI typically predicts outcomes based on available data, often failing to offer the evidence support that clinicians need.

In this paper, we delve deeper into clinicians' hands-on experiences and views on AI-CDSS within the current EHR system.
Focusing on sepsis diagnosis, we aim to design a better form of collaboration between doctors and AI in current medical decision making to provide doctors with better decision support.

\subsection{AI-supported Clinical Decision Support Systems Design}
\label{sub:related_work:design}
There has been growing research on designing AI-CDSS systems based on some general human-AI decision-making research \cite{chiang2023two}.
Researchers design and implement a physician-facing interface and explore how physicians use the system to gain insights.
Typically, AI-CDSS offers decision support to doctors by delivering predictions, risk evaluations, or suggestions \cite{sendak2020human,cai2019human,beede2020human,lee2021human}.
For instance, \citet{yang2019unremarkable} introduce a system that auto-generates slides containing machine prognostics to aid clinicians in decision making.
Some systems offer supplementary evidence or explanations to help doctors understand AI output, thereby enabling trust calibration. Examples include referencing literature \cite{yang2023harnessing}, comparing with prior data \citet{cai2019human}, and bridging knowledge gaps \cite{fogliato2022goes}.
Recently, some systems also support interactive interpretation aids, such as using attention mechanisms for model explanations \cite{cui2020deterrent} and enabling doctors to delve into nuanced concepts \cite{cai2019human}.

Previous research suggests that in sepsis diagnosis, the most important challenge is not to use XAI methods to explain the model outcome, but to use methods that are consistent with the doctor's cognition to establish trust between the doctor and the model~\cite{10.1145/3351095.3372827,sivaraman2023ignore}.
However, many current explainable AI methods cannot meet this goal for sepsis-related AI-CDSS \cite{bedoya2019minimal,downing2019electronic,wang2021explanations}.
In this paper, we further explore the challenges encountered by doctors in cooperation with AI-CDSS during the current early diagnosis of sepsis.
We then design a better system to provide doctors with better diagnostic support.

\section{Formative study: Current Practices and Challenges of AI-assisted Sepsis Diagnosis}
\label{sec:stage1}
As discussed above, there is a research gap in how clinical experts perceive and interact with the AI-based module in their daily diagnosis workflow around sepsis patient cases. To better understand user needs and design challenges, we begin our project with an open-ended semi-structured interview~\cite{longhurst2003semi} with six domain experts as a formative study to gather information on 1) the clinicians' daily practice of sepsis decision making, and 2) the user experience and user needs of AI-based sepsis decision support systems.

\subsection{Method}

We recruited six clinicians who are domain experts and whose daily work involves decision making about sepsis diagnosis and also have used the Epic sepsis module (ESM). 
We used snowball sampling~\cite{goodman1961snowball} to identify and recruit these participants by first reaching out to our colleagues and connections in related fields and then asking them to refer their connections. 
As shown in Table \ref{tab:Interviewees}, all participants are active physicians or nurses working in departments (Intensive Care Unit, Emergency Department, or Internal Medicine) where clinicians most likely encounter sepsis patients. 
In the online pre-screening survey, all participants reported that they ``have used or are still using'' the sepsis decision support module (ESM) in their EHR system (i.e., EPIC system).
All interview sessions were conducted remotely via Zoom, and each interview session lasted, on average, 35 minutes. 
This study was pre-approved by the IRB committee of the first author's institution.

During the interviews, participants were asked to recall a recent case of sepsis encounter, detailing how the diagnosis decision of sepsis was made and what information and factors led to their final diagnostic decision, while not disclosing the patient's personal identifiable information (PII). 
Grounded in that aforementioned sepsis encounter experience, we prompted our participants to report further on their interaction and user experience with existing information technology (IT) systems, such as the EHR and ICU patient monitoring system. 
In particular, we asked them how they interact with and think about AI-driven ESM in their daily diagnostic process. 
We concluded our interview with their attitudes and user needs on the trend of deploying AI (not specific to EMS) to the medical decision-making process.
The detailed semi-structured interview protocol can be found in Appendix~\ref{appendix:interview_script}.

All interview sessions were audio-recorded and transcribed with the interviewees' consent. 
We employ an inductive approach~\cite{thomas2006}, where two researchers in our team first independently coded the interview transcripts, then discussed and reconciled the coding schema, and finally reiterated and re-examined the data with the coding schema.


\begin{table}
    \centering
    \caption{Demographics of Physicians Participants. ICU - Intensive Care Unit, ER - Emergency Room, IM - Internal Medicine. }
    \resizebox{\linewidth}{!}{
    \begin{tabular}{ccccc}
    \toprule
      \textbf{Participant ID} & \textbf{Gender} & \textbf{Departments} & \textbf{Job Title} & \textbf{Year of Experience}\\
    \midrule
      P1  &  Female & ICU      & Staff Nurse & 17 years\\
      P2  &  Male   & IM     & Physician   & 14 years\\
      P3  &  Male   & IM       & Physician   & 14 years\\
      P4  &  Female & ER       & Physician   & 4 years\\
      P5  &  Male   & ICU \& ER & Physician   & 16 years\\
      P6  &  Female & ER       & Physician   & 10 years\\
  \bottomrule
    \end{tabular}
    }
    \label{tab:Interviewees}
\end{table}

\subsection{Result}\label{sec:result}

We found that clinicians' decision-making process for sepsis diagnosis is high-stakes (life-threatening), time-sensitive (a patient's severe symptoms developed only in a few hours), and very uncertain (need extensive lab test outcomes that they often do not have at the moment of decision-making).

\begin{figure*}
    \centering
    \includegraphics[width=0.95\linewidth]{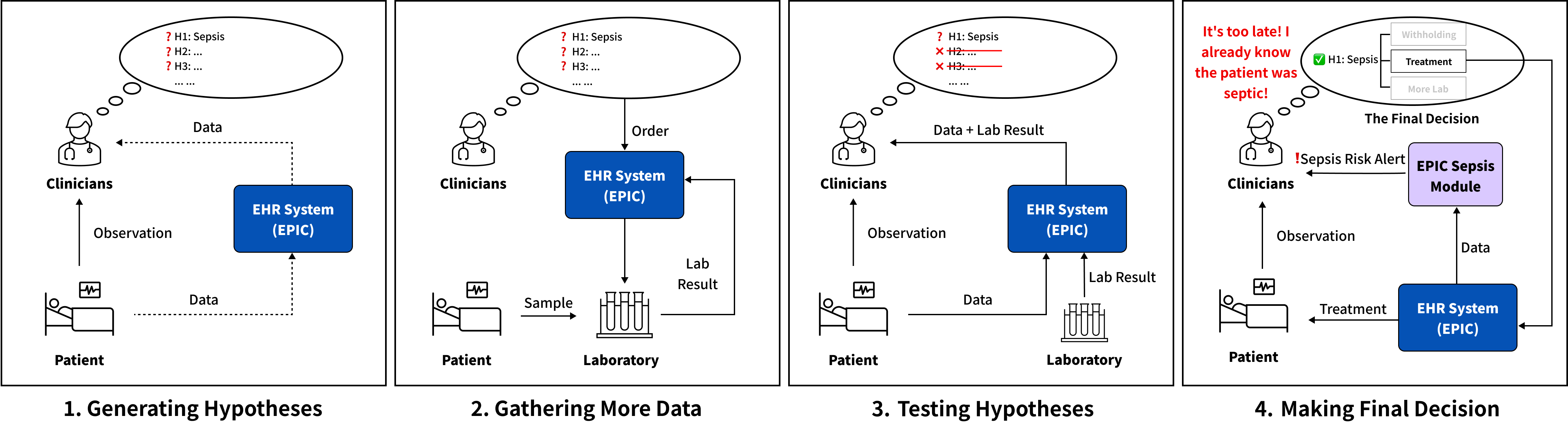}
    \caption{\textbf{Existing Human-AI Interaction and ``Competition'' Paradigm.} The current sepsis module mainly focuses on supporting the final decision-making stage \cite{sox2013medical}, yet physicians often find the AI predictions are too late and not helpful.}
    \label{fig:existing-hai-collab}
\end{figure*}

Clinicians walked us through the procedure of sepsis diagnosis, which provided a multifaceted journey from the patient's entry into the ER and (potentially) moving to the ICU or IM. They offered us more insights to enrich the $4$-step procedure in Section~\ref{sub:related_work:sepsis}.
We summarized the workflow in Figure~\ref{fig:existing-hai-collab}:
(1) \textit{Generating Hypotheses}. For a patient who has the risk of sepsis, physicians access and read a patient's vital signs and current state from the EHR system. They form a set of hypotheses given the patient's situation, yet these hypotheses are unclear with the large uncertainty.
(2) \textit{Gathering Information.} Based on the most promising hypotheses, they order specific lab tests to collect more information related to these hypotheses.
(3) \textit{Testing Hypotheses}. According to the lab test results, physicians narrow down or scale up their hypotheses.
(4) \textit{Making Action Decisions.} Based on the new hypotheses, physicians make decisions among three options: treating the patient, gathering more information, or withholding and waiting for new development of the disease.

All participants in general welcome the future of having more AI support for medical decision making, but, to our surprise, they strongly believe that the current Sepsis Module in EPIC (ESM) is not only useless, but also leads to additional meaningless work.
Participants perceived the current ESM as a simple \textit{sepsis risk prediction score}, and their dissatisfaction comes from (a) the risk prediction score is often \textbf{belated} (Section~\ref{sub:stage1:belated}), (b) \textbf{inaccurate} (Section~\ref{sub:stage1:inaccurate}), (c) \textbf{no explanation} (Section~\ref{sub:stage1:explanations}), (d) \textbf{no actionable insights} (Section~\ref{sub:stage1:actionable_insights}).
In addition to these surface-level concerns regarding system design and algorithm performance, participants believe that the most fundamental issue is that the current paradigm of human-AI interaction design has a (e)\textbf{wrong focus} of AI assistance --- it attempts to support the final decision of a complex medical decision-making process (i.e., high-stakes, time-sensitive and high-uncertainty). Together with the other issues mentioned above, human decision-makers feel challenged or even intimidated by the AI system, which eventually leads to all participants totally ignoring the current sepsis AI module (Section~\ref{sub:stage1:threat}). 
We organize the results with these five aspects and dive into each of these issues.

\subsubsection{Belated Sepsis Risk Prediction}
\label{sub:stage1:belated}
All physicians complain that the current sepsis prediction is too late and thus useless in their decision-making process.
This is due to the fact that sepsis is a life-threatening disease and can progress very fast. Thus, sepsis diagnosis requires human decision-makers to make a time-sensitive decision at the first encounter with the patients, despite they face huge uncertainty due to the lack of sufficient information about the patients.
However, AI prediction must rely on data as input, but at first encounter, many of such data (e.g., vitals or lab results) do not exist or are not digitalized.
\begin{quote}
  \textit{``So the big thing is that if we went strictly by the tool [AI-predicted sepsis cases] for our ED patients, we would be usually like three to four hours behind [human-diagnosed cases].''} (P6)
\end{quote}

An extreme but quite common case reported by participants (P4, P6) is that shortly after human clinicians made the decision that a patient has sepsis, recorded the decision in EHR, and started to put in laboratory orders and antibiotic treatments in the EHR, the current sepsis risk prediction module consequently predicted that this is a sepsis case (see Figure~\ref{fig:existing-hai-collab} Step 4).
Even worse, the current EHR system and hospital policy mandates the nurse or clinician to respond to this sepsis alert.
\begin{quote}
  \textit{``If it triggers after two or three hours, I already know that, and I've already been treating them.''} (P6)
\end{quote}

Most participants explicitly demand an \textbf{early prediction} for early diagnosis of sepsis (P3, P4, P6), most likely at the first encounter in the emergency room (ER).
They believe that such AI prediction could significantly speed up the following procedures and improve patient's final outcomes.
However, the current sepsis prediction AI model cannot achieve that.

\subsubsection{Inaccurate Sepsis Risk Prediction}
\label{sub:stage1:inaccurate}

The current algorithm design tends to have an extremely low sensitivity threshold at 13\% (i.e., high false positive rate at 87\%) to avoid missing any potential sepsis, because sepsis decision making has a patient's life at stake.
\begin{quote}
    \textit{``the majority of times, I think it's inaccurate''} (P3)
\end{quote}
As a result, participants (P3, P6) reported that they received an overly large volume of false alerts from the sepsis module due to the inaccurate sepsis risk prediction algorithm and the low sensitivity threshold.

\begin{quote}
    \textit{``Any system that produces tons of alerts will induce alert fatigue, and people won't pay attention to it ... on average, there are already more than 20 interruptions per hour for an ER physician. So if you are adding more [inaccurate] interruptions to me, I'm not gonna pay attention, it becomes noise in the background.''} (P6)
\end{quote}

Even worse, due to the severe consequences of sepsis, the EHR system and hospital policy \textbf{forcefully mandated clinicians to manually verify} if they had taken appropriate diagnostic or treatment actions, or dismiss the alert but with a mandate note to explain to the system why the human clinician did not take any action (e.g., they have already taken sepsis treatment actions before the alert).
\begin{quote}
    \textit{``The nurses get the BPA (sepsis risk alert) fired when it triggers that score. And then [nurses] have to put in [some notes] like an acknowledgment. A couple of options are like, ``treatments already initiated'', or ``notified physician''. Or you can silent it, I think, for 15 minutes or for half an hour, and then it'll fire again.''}
\end{quote}

\subsubsection{Lack of Explanations}
\label{sub:stage1:explanations}

Participants (P1, P4, P5, P6) also mentioned that the current AI-predicted sepsis risk score is hard to interpret.
Participants do not understand why an obvious sepsis case has a lower risk score than the score of a less obvious case.
\begin{quote}
    \textit{``I don't think [another AI with higher prediction performance] would change anything because we were already unsure [how the current one works], and we were already looking for more explanations''} (P4)
\end{quote}
They found that their multiple years of medical experience could not help interpret the relativeness of the score or the factors contributing to a score.
\begin{quote}
    \textit{``On the Epic, there are too many parameters that I don't remember. [But] I know that it does not follow any of the other sepsis diagnosis criteria [being taught and used in practice] with the one, two, or three rating scale. And I know that it can go really high. And there are too many parameters.''} (P3)
\end{quote}

Interestingly, the designer of the system already incorporated a feature importance score (a percentage of how much each factor contributes to the final prediction) as a simple explanation of the AI prediction, but they were hidden too deep in the interface. As a result, none of the participants except one (P4) was aware of its existence.
\begin{quote}
    \textit{`` I honestly never looked that deep into it (the risk score), I just see the color [of the risk score], and then I just go through my [``human''] algorithm that I've done for years and years.''} (P2)
\end{quote}
Even if such feature important percentage scores are at the top level, participants may still ignore them as there is already too much numerical information on the EPIC EHR interface.
\begin{quote}
    \textit{``The good thing about EPIC and the worst thing about EPIC are the same --- everything is there. It's kind of hard to figure out what you need to know. And it sometimes takes too many clicks.''} (P3)
\end{quote}

\subsubsection{No Actionable Insights}
\label{sub:stage1:actionable_insights}

Our interviewees also questioned the limited utility of it in their medical diagnostic process of sepsis.
As we mentioned earlier, participants generally followed the four steps of medical decision making, from formulating hypotheses in their minds to finally making actionable diagnostic or treatment decisions. 
Most clinicians (P1, P2, P3, P4, P6) mentioned that they simply ignored the risk score during their diagnosis process because they were confused about the purpose of the risk score and did not know what to do as an actionable next step given a high or low risk score.
\begin{quote}
\textit{``So I'm not really sure what its goal is, but I can tell you that most of us ignore it (the sepsis risk score), because it has not proved helpful to what we do next''. }(P4)
\end{quote}

\subsubsection{AI Helper or AI Challenger?}
\label{sub:stage1:threat}
The issues raised by participants in (Section~\ref{sub:stage1:belated}, Section~\ref{sub:stage1:inaccurate} and Section \ref{sub:stage1:explanations} may be addressed with a better algorithm or a more advanced user interface design. 
However, the concern of \textit{lacking actionable insights }(Section~\ref{sub:stage1:actionable_insights}) hinted at a more fundamental challenge that goes beyond the algorithm and interface design ---
the current AI-based sepsis prediction module mainly focuses on the prediction of the final decision outcomes, in our case, it is the sepsis risk score that suggests whether a patient has sepsis.
Such output at the final decision stage implicitly ``challenges'' human decision-makers' authority and expertise in their roles of making that final decision.
\begin{quote}
    \textit{``[The AI module] tries to just make a decision and tells me that [these patients] might have sepsis, so I have to do everything I'm supposed to do for [treating] them. That doesn't help.''} (P5)
\end{quote}

Due to the current prediction focusing too much on the final stage of decision making only, this human-AI interaction paradigm is essentially perceived as a human-AI competition, which challenges the human experts' expertise and intimidates their authorities and feelings.

\begin{quote}
    \textit{``I think that AI can be very helpful as part of patient care, but I don't think it should replace the care and decisions a physician can make''} (P5)
\end{quote}

\begin{figure*}[t]
    \centering
    \includegraphics[width=0.95\linewidth]{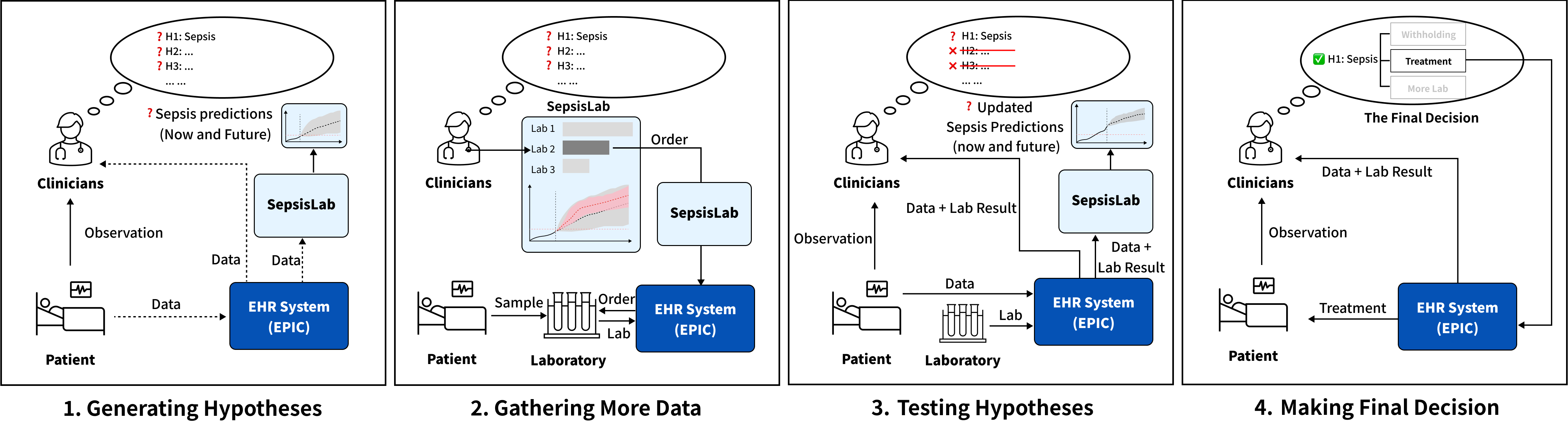}
    \caption{\textbf{The Clinician's Medical Decision-Making Workflow with Support from \system.} \system focuses on providing support to the intermediate steps of the clinical experts' decision-making process \cite{sox2013medical}, as opposed to existing AI modules that focus only on the final decision-making stage.
    \system can generate predictions for the patient's sepsis onset possibility (as the risk score) now and in the future (\textbf{\textcolor{purple}{Design Strategy 1}}, \textbf{\textcolor{orange}{Design Strategy 4}}), as shown in Step 1; It can further suggest additional lab tests by their impact on model uncertainty (\textbf{\textcolor{blue}{Design Strategy 2}}), and the interactive visualization can help clinicians select the most valuable lab tests to support their decision (\textbf{\textcolor{DarkGreen}{Design Strategy 3}}, \textbf{\textcolor{orange}{Design Strategy 4}}), as shown in Step 2; Once new data are collected, the prediction visualization will be updated (Step 3), helping clinicians test hypotheses. Then, following our \textbf{\textcolor{magenta}{Design Strategy 5}}, clinicians can generate new hypotheses or reach final decisions (Step 4).}
    \label{fig:system}
\end{figure*}

Instead, participants believe that \textbf{AI can assist human experts in other places or stages of the medical diagnosis process}. For example, it can simply propose the sepsis possibility as a candidate hypothesis in the medical decision-making process.
\begin{quote}
    \textit{``I would say that whether [AI] gave me like a 10\% or a 90\% sepsis risk score, I'm not sure that that would change my [decision]. If it [AI] simply tells me to think about it [sepsis possibility], then I'll just go to think about it.''}
\end{quote}
Alternatively, AI can suggest what kinds of laboratory data can be collected to support the test of the candidate hypothesis, from which physicians can obtain their desired actionable insights, and the uncertainty level can be reduced.

\begin{quote}
    \textit{``We want to be better at knowing what to order and when and how to order it [lab or treatment]... in the current way that we're pushed by [the AI sepsis prediction score] right now is not useful, [and] it is correct most of times ... If a predictive model can trigger us to take an action [such as ordering lab or treatment] that prevents patients from getting sicker. That'd be amazing.''} (P4)
\end{quote}

\subsection{Summary of Results}
In summary, our formative study shows that the existing AI-driven sepsis risk prediction module does not support clinicians in their medical decision-making scenarios, because
the current sepsis prediction algorithm is belated and inaccurate, the interface does not have explanations, and the AI prediction cannot be transformed into a diagnostic or treatment action.
These challenges reveal a fundamental issue of the existing human-AI decision-making paradigm that human experts need AI to focus more on supporting their intermediate decision-making process, rather than predicting a final outcome.
These findings shed light on our design of a new sepsis module with the goal of a new human-AI collaboration paradigm.

\section{\system: a Human-Centered AI System to Support Early Diagnosis of Sepsis}
\label{sec:system}

In this section, we will start with the design strategies derived from the results of the formative study. Then, we will present both the user interface and the back-end algorithm of a novel human-centered AI system, \system, which aims to implement those design strategies to support clinical experts in making diagnostic decisions about sepsis. 
Our \system can predict the patient's current sepsis risk score, as well as the sepsis risk in the next 4 hours, based on patient history information and available vital signs and lab test values. 
Often times, some lab test results are missing but may also be critical for the diagnosis of sepsis. Therefore, our system can rank the top $5$ lab tests that can reduce the uncertainty of the prediction and show them as recommendations to clinicians.
Furthermore, our system has a counterfactual prediction module so that users can interactively review how each missing lab result may improve the prediction or reduce uncertainty before actually performing this lab test. 

\subsection{Design Strategies} 
\label{sub:system:design}

Based on the stage 1 findings, we conclude five design strategies for the new design of the sepsis module.

\textbf{\textcolor{purple}{Design Strategy 1}: Performing Future Risk Score Prediction.}
As our formative study results suggest, clinicians do not need an inaccurate and even belated risk score prediction (Section~\ref{sub:stage1:belated} and \ref{sub:stage1:inaccurate}). Instead, they need an accurate prediction score that is predicted ahead of time. This is also supported by previous empirical studies in sepsis diagnosis~\cite{10.1001/jamainternmed.2021.3333,wong2021external,silvestri2022desired}, which
requires a better algorithm in the back-end of our system.

\textbf{\textcolor{blue}{Design Strategy 2}: Providing Accessible Model Explanation.}
Our interview results also suggest the need for an easily accessible section for sepsis risk prediction explanations (Section~\ref{sub:stage1:explanations}).
This is a common issue found by previous research~\cite{10.1145/3351095.3372827}.
\system needs to have a simple design to present explanations in an easy-to-find and easy-to-understand manner.

\begin{figure*}[t]
    \centering
    \includegraphics[width=0.9\linewidth]{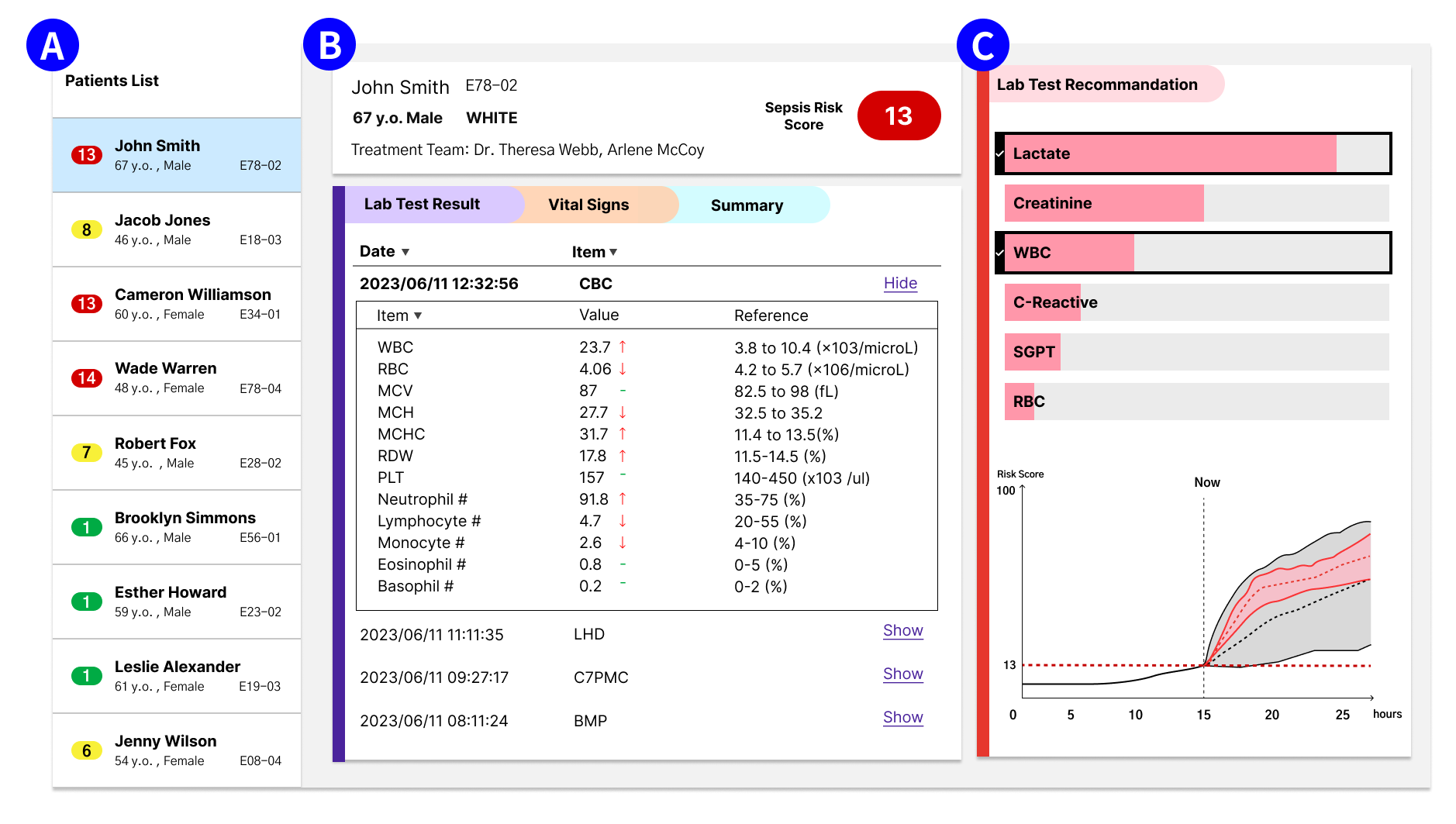}
    \caption{\textbf{User Interface of Our Prototype System. }
    (A) A list of patients with different sepsis risk prediction scores, colored from no risk as Green, to medium risk as Yellow, to high risk as Red.
    (B) The patient's demographics and the dashboard that includes the patient's vital signs, lab test results, and medical history.
    (C) Our \system system as an add-on to the existing EHR system.
    This UI currently illustrates that a clinical expert is examining a high-risk patient's data who was admitted 15 hours ago.
    The AI suggests the expert collect more lab results. The expert is interacting with the visualization to see if Lactate and WBC lab results were added, how the sepsis prediction and its uncertainty would change. 
    All patient names and demographic information in this screen capture are random generated fake data for illustration purposes.
    }
    \label{fig:UI}
\end{figure*}

\textbf{\textcolor{DarkGreen}{Design Strategy 3}: Revealing Actionable Insights and Suggestions.}
Moreover, a risk score, even predicted for future timestamps, cannot provide actionable insights, as suggested in Section~\ref{sub:stage1:actionable_insights}.
Clinicians base their diagnostic decisions on physical signs and lab test values from the patient, where gathering data (i.e., lab tests) plays an important role (recall Figure~\ref{fig:human-AI-collab-paradigm}).
Therefore, \system is designed to generate meaningful recommendations about potential lab tests.

\textbf{\textcolor{orange}{Design Strategy 4}: Displaying Uncertainty beyond the Risk Score.}
Sepsis diagnosis is a highly uncertain decision-making process. Providing a single risk score value may miss important information. Therefore, \system also calculates and displays the uncertainty in addition to the risk score. This is also aligned with the previous work about the advantages and benefits of XAI \cite{wang2021explanations, zhang2020effect}.

\textbf{\textcolor{magenta}{Design Strategy 5}: Shifting AI from Suggesting the Final Decision to Supporting Intermediate Stages.}
Finally and most importantly, the key takeaway from our formative study suggests the need to shift AI's focus. Existing AI-based sepsis module mainly focuses on the final decision stage, creating a sense of competition for physicians and leading to the abortion of the module (Section~\ref{sub:stage1:threat}).
To address this challenge, \system is designed to support human experts' intermediate decision making stages, including generating hypotheses, gathering data, and testing hypotheses (Step 1 to 3 in Figure~\ref{fig:human-AI-collab-paradigm}).
In such a way, our system can build a new human-AI collaboration paradigm, where AI can actually team with experts to support what they need.

Combining these five design strategies, \system supports a new medical decision-making workflow for sepsis diagnosis, as shown in Figure~\ref{fig:system}.
\system can generate predictions for a patient's sepsis risk score now and in the future, which can support the generating hypotheses stage.
Moreover, it can further suggest additional lab tests that clinicians may gather to support their decisions. With the interactive visualization, our system helps them select the most valuable lab tests. This can provide actionable insights and support clinicians' data-gathering process.
Once new lab test data are collected, the prediction visualization will be updated and assist clinicians in testing their hypotheses.

\subsection{Front-End User Interface Design}\label{sec:systemoverview}

Following the design strategies, we designed and implemented the new sepsis module based on human-AI interaction guidelines \cite{amershi2019guidelines, shneiderman2010designing}.
The user interface (Figure \ref{fig:UI} \footnote{Note that we used MIMIC-III data for our algorithm illustration. All patient names and demographic information in this screenshot are random generated fake data for illustration purposes.}) includes three parts:  (A) Left: The current patient list,
(B) Middle: The selected patient’s demographic information, medical history, lab test results, and vital signs monitoring,
(C) Right: Our AI-powered Lab Test Recommendation Module, \system, including Lab test recommendation, risk score predictions, and counterfactual explanation.
We used de-identified patient data from MIMIC-III \cite{johnson2016mimic} as the data of the prototype system (data in Figure \ref{fig:UI} B \& C). 

It is noteworthy that we deliberately designed the sepsis score of our predictive model's outcome to be the same as that of the current EPIC Sepsis Module. So that we could help \system have an easy integration into the existing EPIC system, where the clinicians can find it familiar and easy to use.
We introduce each feature in Figure \ref{fig:UI} C one by one.

\textbf{Future Risk Prediction with Uncertainty Visualization.}
As mentioned in Section~\ref{sub:system:design}, one core part of \system is a predictive algorithm that generates future prediction of sepsis risk scores (\textbf{\textcolor{purple}{Design Strategy 1}}).
As shown in the bottom part of \system interface, we design a time-series plot to visualize both the historical (the solid line) and expected future (the dashed line) risk prediction trajectory over time.

Moreover, for each risk score prediction, the model also generates the uncertainty range of the expected value, as shown in the gray area in the line plot (\textbf{\textcolor{orange}{Design Strategy 4}}).
We selected visualized confidence intervals to display the prediction uncertainty, because prior work has found that confidence intervals evoke high levels of trust~\cite{padilla2022, shneiderman2010designing}. 
While prior research has found that other uncertainty visualization techniques produce more accurate risk judgments~\cite{padilla2021}, our visualization aims to show relative risk rather than for clinicians to read off specific values. Future work would be apt to consider the effects of various visualization design choices.

\textbf{Feature Importance Visualization.}
The algorithm takes lab test item values as the input and generates risk score prediction.
Each newly collected lab test item can be used to update the model and (potentially) reduce the model uncertainty.
Therefore, we designed a ranked horizontal barplot on the top of the \system interface to visualize the important items that contribute to the prediction uncertainty reduction (\textbf{\textcolor{blue}{Design Strategy 2}}).
The item with the highest importance is ranked on top of the barplot.

\textbf{Lab Tests Recommendations.}
Combining the two parts of future risk prediction and feature important visualization, we added the lab tests recommendation function into our interface (\textbf{\textcolor{DarkGreen}{Design Strategy 3}}).
As mentioned above, different newly collected lab items can change the model prediction uncertainty.
Therefore, \system recommends an item list ranked by their importance. The clinician can select one or multiple lab items and observe how their test results could influence the model's risk prediction trajectory (the red dashed line) and the corresponding uncertainty range (the red area).
Note that the red line and area are counterfactual values that are estimated by the algorithm (more details in Section~\ref{sec:labtestdesign}).

\begin{figure*}[!t]
\centering
\subfloat[ ]{\includegraphics[width=0.30\linewidth]{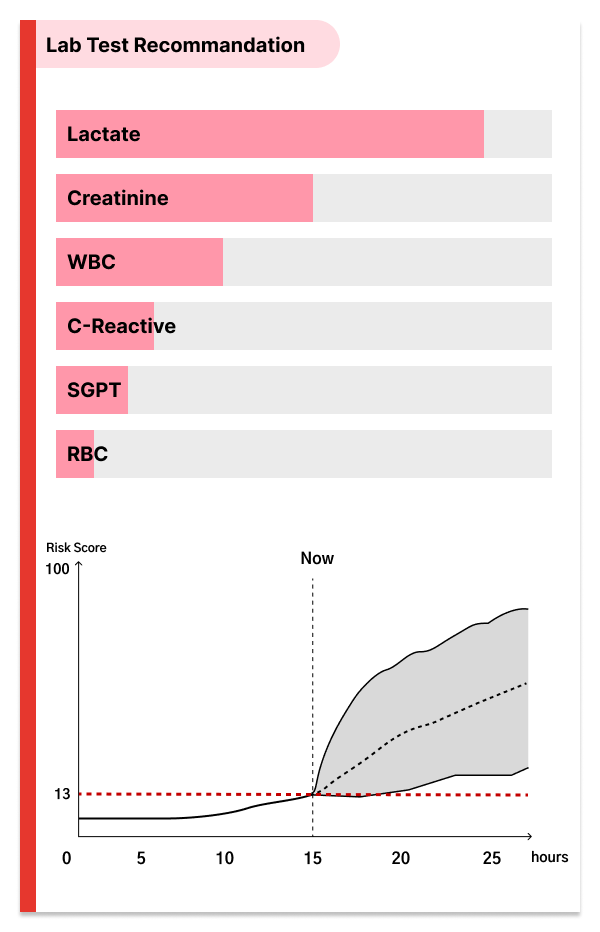}
\label{fig:ltr1}}
\hfil
\subfloat[]{\includegraphics[width=0.30\linewidth]{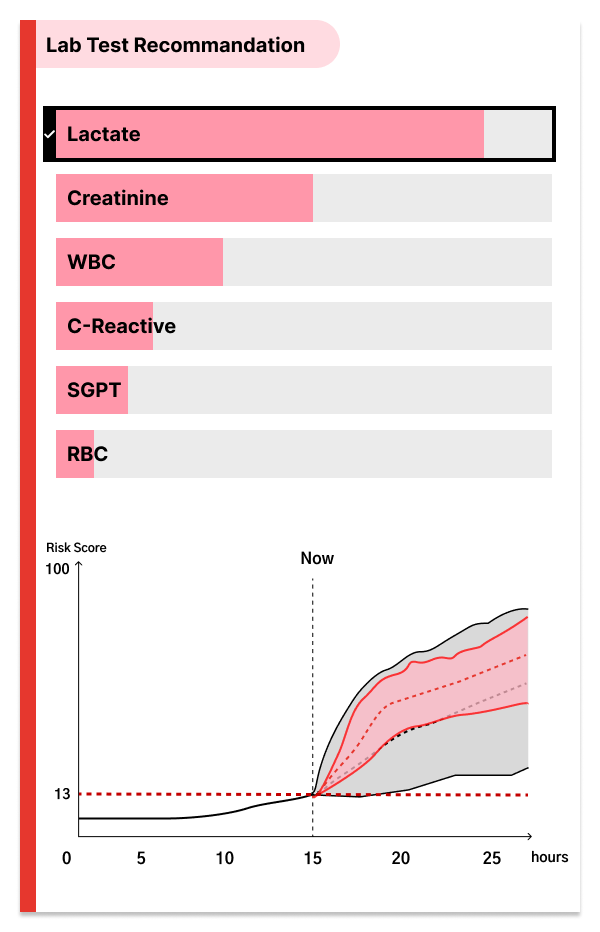}
\label{fig:ltr2}}
\hfil
\subfloat[]{\includegraphics[width=0.30\linewidth]{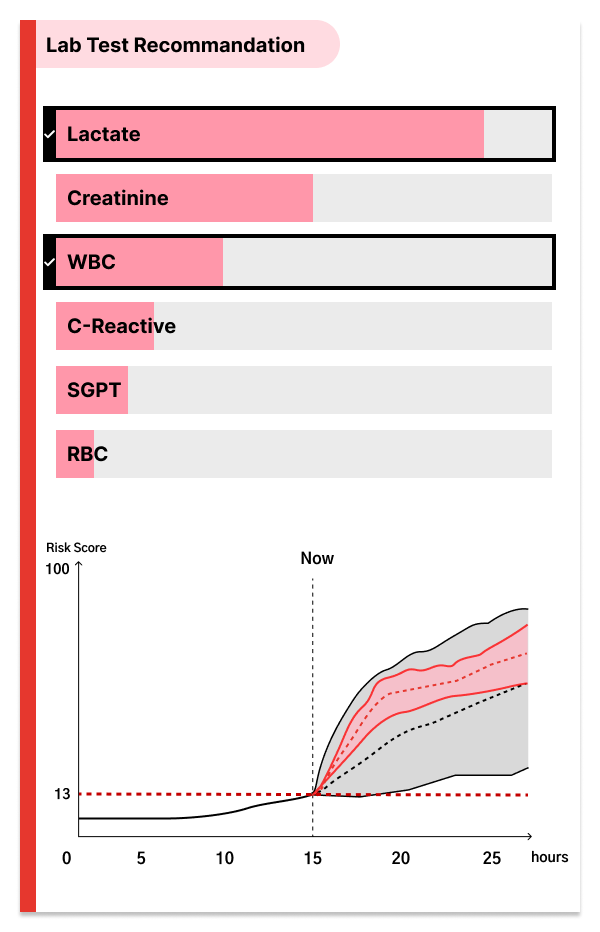}
\label{fig:ltr3}}
\caption{\textbf{The Interactive Lab Test Recommendation Module in \system.} (a) The clinician can get an actionable lab item test recommendation list from \system. The items are ranked by their importance to reduce the uncertainty of the sepsis future prediction. (b) The clinician can interact with \system to select a lab item and see its expected influence of the lab test result on the model uncertainty via a counterfactual prediction. (c) The clinician can select multiple lab items and see their combined expected influence of the results on the uncertainty.
}
\label{fig:ltr}
\end{figure*}

\system supports the clinicians to interact with the interface. Figure~\ref{fig:ltr} visualizes the interactive process by picking different potential lab test items.
By comparing different combinations of the lab test items, the clinician can obtain a better understanding of the model and make the decision to order appropriate lab tests to collect the actual item values, which then truly update the model's prediction trajectory and uncertainty range.
Overall, the interface follows Section~\ref{sub:system:design} to support clinicians' intermediate decision making stages (\textbf{\textcolor{magenta}{Design Strategy 5}}).

\subsection{Back-End Algorithms Design}\label{sec:labtestdesign}

To support the design strategies (Section~\ref{sub:system:design}) and the UI features (Section~\ref{sec:systemoverview}) informed by the formative study results, our back-end consists of three sub-modules: 1) an LSTM-based predictive model that can take only partial or little data of a patient as input and generate prediction scores of sepsis risk for the patient in the upcoming period as output; 2) a lab test recommendation module based on the uncertainty estimation from the previous predictive model; and, 3) a counterfactual generation module that can show the users how a hypothetical lab result may change the sepsis risk prediction scores and uncertainty range of the predictive model, and enables users to interact with the visualization chart in the front-end.

\subsubsection{LSTM-based Predictive Model for Sepsis Risk Prediction}\label{sec:prediction_algo}

EHR data are typically a temporal sequence of patient activities in a hospital system. Depending on how frequently a patient visits a doctor or takes a lab test, the data sequence may be very sparse and irregular for a particular patient. 
Prior works \cite{choi2016retain,choi2017gram,ma2018kame,yin2019domain,zhang2021interpretable} suggests Long Short-Term Memory (LSTM)~\cite{hochreiter1997long}, a special type of RNN model, has consistently demonstrated their remarkable performance in clinical risk prediction tasks using EHR data (see Table~\ref{tab:classification_results} from ~\cite{zhang2021interpretable} in Appendix \ref{app:LSTMperformace} as the evidence of the model performance).
Recurrent Neural Network (RNN) model architectures are more suitable for the sepsis predictive tasks on the temporal observational data with irregular time intervals. In this study, we select Long Short-Term Memory (LSTM)~\cite{hochreiter1997long} as the backbone of the prediction framework, which is able to capture both long-term and short-term clinical information in patients' EHR history, and thus improve clinical prediction performance.

To satisfy design strategy (1) that clinical experts want a prediction model that has a high accuracy and can predict ahead of time, we adopt the LSTM-based sepsis prediction model from~\cite{zhang2021interpretable} as our base model to support the prediction of the sepsis onset risk in the next 4 hours.
We provide our LSTM model implementation details and parameters in Appendix \ref{app:modelpara}.
The extracted static information vector (e.g., patient's demographic and history) is used to initialize the hidden state of LSTM. 
Then, the LSTM takes a sequence of collection data (e.g., vital signs and laboratory values) in addition to their occurring time as inputs and generates a sequence of the latent health state. 
Sometimes an observation may be missing (e.g., a patient has not performed a lab test or their previous lab test result has been outdated), thus a value embedding \cite{yin2020identifying} is used to map the observations into vectors.

A variable attention module that can handle varying numbers of inputs is followed to generate a fixed-size vector that is sent to LSTM. 
With such a model design, the predictive model can start making predictions at the first encounter with the patient even there is not much data, which satisfies the \textbf{\textcolor{purple}{Design Strategy 1}} that users want to see predictions into the future.
The attention module can automatically focus on important variables, and the learned attention weights can be used to interpret the prediction results --- this enables the users to see each input feature's importance score contributing to the predictive model --- satisfying \textbf{\textcolor{blue}{Design Strategy 2}}. 
After all the output vectors of LSTM are produced, a collection attention module is followed to combine the sequence of output vectors into a vector. 
Finally, a fully connected layer and a Sigmoid layer are followed to predict the sepsis onset probability.

\subsubsection{Lab Test Recommendation Based on Uncertainty Estimation} 
\label{uncertainty_estimation}
The AI model's prediction always comes with certain degrees of uncertainty. In the sepsis early diagnosis scenario, a new patient when they just arrived ER may not have any lab test results in the EHR, thus many missing values as input to the predictive model.
Due to this uncertainty, simply looking at the predicted sepsis risk score without the certainty level, users may not be able to accurately evaluate the trustworthiness of an AI prediction, and that is why participants reported that a patient with a high risk score for sepsis is not necessarily more accurate or more urgent than a patient with a low risk score. 

We estimate uncertainty and reduce uncertainty via improving the aforementioned LSTM-based predictive model:
We hypothesize that the missing variables follow a Gaussian distribution so that we can estimate the parameters (i.e., mean and covariance) for each missing variable. 
Based on previous work that has shown superior performance in missing value imputation with deep learning \cite{yin2020identifying},  we adopt Monte Carlo Simulation (MCS) to sample the missing values many times and compute the uncertainty with the standard deviation of the outputs with MCS. 

Two thresholds $th_s$ and $th_e$ ($1>th_s, th_e>0$) are set according to the desired sensitivity and precision to decide whether new laboratory values should be requested. 
If the output sepsis onset probability $p>th_s$, the model predicts that the patient will have sepsis onset after 4 hours. 
Sometimes, the model may be uncertain about the prediction results. We define the \textbf{uncertainty} as Shannon entropy \cite{shannon1949mathematical}:  
$ e = - p \log (p) - (1-p)\log(1-p)$. 
If the uncertainty $e>th_e$, the model will \textbf{recommend clinicians to collect more clinical variables}, for example laboratory values. 
Then, the model takes the updated values as input and can output new results with higher confidence.

To confirm that the recommendation can improve the model performance in the absence of certain laboratory values, we conducted experimental tests on the MIMIC-III dataset \cite{johnson2016mimic}.
As shown in Table \ref{tab:active_sensing_results}, in our implementation, with an LSTM model, our recommendation algorithm performs comparable to full observation setting models and outperforms the masked setting models by approximately 10\% with only 9.6\% extra laboratory values requested by our recommendation algorithm.
The results indicate that our recommendation algorithm can achieve performance nearly equivalent to that under full observation, thus enabling accurate predictions even with fewer lab test results, without compromising prediction precision.
We provide our recommendation algorithm implementation details including model parameters in Appendix \ref{app:modelpara}.

\begin{table*}[t]
    \centering
    \caption{Improvement of Our Recommendation Algorithm on AUC on MIMIC-III. Masked: all lab test results are deleted (simulating a patient just arriving at the ER). Our Algorithm: 9.6\% lab test results are actively selected and repeated (simulating the clinician carefully selected most critical lab tests to order for the patient and gradually adding more labs if necessary). Full-observed: all the lab test results are used for prediction (assuming a patient has been hospitalized for a long time, so they have done many labs). The results show that our recommendation algorithm performs comparable to full observation models and outperforms the masked setting models by about 10\% with only 9.6\% extra laboratory values requested. The improved column denotes the improved performance with our recommendation algorithm, compared to the results in the masked setting.}
    \begin{tabular}{lccc}
        \toprule
        \textbf{Method} & \textbf{Masked} & \textbf{Our Recommendation Algorithm} & \textbf{Full-observed} \\ \midrule
        Logistic regression & 0.72 & 0.78 & 0.79 \\
        Random forest & 0.75 & 0.81 & 0.83 \\
        Gradient boosting trees & 0.75 & 0.83 & 0.85 \\ \midrule 
        LSTM \cite{zhang2021interpretable} & \textbf{0.79} & \textbf{0.89} & \textbf{0.90} \\
        \bottomrule
    \end{tabular}

    \label{tab:active_sensing_results}
\end{table*}

This algorithm design satisfies both \textbf{\textcolor{DarkGreen}{Design Strategy 3}} and \textbf{\textcolor{orange}{Design Strategy 4}}. The model focuses on recommending missing laboratory values to the clinician so that the clinician perceives such recommendations as an actionable suggestion. Additionally, the uncertainty estimation and visualization shift users' attention from the accuracy of the AI-predicted final decision, but to the reduction of such uncertainty in decision making. 

\subsubsection{Counterfactual Prediction to Explore Uncertainty Reduction Without the Cost of Performing a Lab Test.}

From AI's perspective, it would love clinicians to perform all kinds of lab tests on a patient so that it can reduce most of its uncertainty and predict at a higher accuracy.
However, from the human's perspective, it is too costly and inhumane.
While the lab test recommendation algorithm can identify the most informative laboratory test that is missing and reduce uncertainty of the prediction, it's still hard for doctors to intuitively see the value of these labtests.
Therefore, to further help doctors make decisions we design a mechanism that the algorithm can also output counterfactual predictions to show \textbf{how much uncertainty can be reduced} without actually performing and collecting the recommended laboratory test results, which can provide more information in the process.
For each variable i, we first sample the possible values $K$ times. For each sampled value $x_{i,k}$, we adopt MCS to sample the missing values and compute the uncertainty with the standard deviation of the outputs (denoted as $U_{i,k}$). 
If the variable $i$ is observed, the new uncertainty would be $U_i = \frac{1}{K}\sum_{k=1}^K U_{i,k}$. 
We select the variables with the maximal uncertainty reduction: $i^* = \arg \max_i U - U_i = \arg \min_i U_i$. We set $k=500$ in our implementation. 
The expected uncertainty decrease is $U - U_i$ for variable $i$.
With such a model design, the front-end UI can support interactive visualization that allows users to explore the different laboratory test's effectiveness and further satisfies \textbf{\textcolor{magenta}{Design Strategy 5}}.

\subsection{System Implementation} 

The interactive front-end user interface (Section~\ref{sec:systemoverview}) is developed as a web application using the React framework. 
The particular visualization that enables users to interactively explore the laboratory test recommendations (Section~\ref{sec:labtestdesign}) is developed using the.  Recharts~\footnote{\url{https://recharts.org/}} library. 
We developed the back-end with Python's FastAPI \footnote{\url{https://fastapi.tiangolo.com/}} library. 
The predictive model and the counterfactual model inside the back-end (Section~\ref{sec:prediction_algo}) are implemented with PyTorch \cite{paszke2019pytorch}. 
We primarily store the data (i.e., MIMIC-III \cite{johnson2016mimic}) in a master-slave backup MySQL database for efficient querying and security purposes.
The entire prototype system (front-end, back-end, model, and database) is hosted on Amazon Web Service (AWS) server instances.

\section{Evaluation Study: Enhancing Human-AI Collaboration}
\label{sec:stage2}
Our system aims to implement a human-AI collaboration paradigm for the decision-making scenario for sepsis diagnosis. 
To do so, we design our \system system that shifts the focus of the AI predictions to offering human experts their desired and actionable recommendations in the intermediate stages of a medical decision-making workflow. 
In this section, we report on a heuristic evaluation study by inviting the same six clinicians to interact with and provide feedback on our \system design.
Our findings demonstrate that \system is generally appreciated as it can provide meaningful assistance to clinicians. The participants would love to see such a system deployed into the real EPIC EHR system, and they argue that our AI design can help them achieve a better human-AI team experience.

\subsection{Design and Procedure}

We recruited the same six clinical experts in Section~\ref{sec:stage1} to perform a heuristic test with our system.
We first pre-loaded de-identified patient data from MIMIC-III \cite{johnson2016mimic} into our prototype and deployed it in an internal cloud cluster to prevent leakage of patient data.
To demonstrate the system across varied patient conditions, we selected two patient groups from MIMIC-III \cite{johnson2016mimic}, one with patients ultimately diagnosed with sepsis and the other without sepsis.
Patients ultimately diagnosed with sepsis were selected based on the sepsis-3 \cite{singer2016third} criteria and are all adults.
Limited by the duration of each study session, we randomly selected five data points from each patient group, totaling ten data points for display and interaction with participants.
Due to the usage regulations that researchers need approvals to access for MIMIC-III \cite{johnson2016mimic}, we only present the mock data in the paper to present the UI.
Our \system system was used as a design probe to solicit participants' user experience and design requirements. Participants navigated the interface and thought aloud in testing its different functionalities. We then conducted a semi-structured post-study interview.
Specifically, we asked about their comments on the AI assistance's new focus on intermediate decision-making steps, and whether the lab test recommendations and the counterfactual predictions were practical and could enhance their decision-making process. We also collected clinicians' suggestions for improvements to the system.
Each user study session lasted around 30 minutes. We recorded the interview audio and used an inductive process~\cite{thomas2006} to analyze the interview transcription.

\subsection{Result}
Overall, the participants gave very positive comments on our system prototype. 
With AI shifting focus from final-stage prediction to intermediate stages, clinicians found AI less intimidating but more cooperative.
Meanwhile, participants also liked our new functions that went beyond risk score prediction and commented that they were helpful in revealing more information, providing actionable insights and improving AI transparency.

\subsubsection{Shifting AI Focus Away from Final Decision Prediction Can Enhance Human-AI Collaboration Experience}
As revealed from our formative study, the fundamental challenge of the existing sepsis module lies in the fact that it only focuses on the final stage of the four-step diagnosis process (see Section~\ref{sub:stage1:threat}). 
Our system aims to address this challenge by shifting AI's focus from the last stage to the intermediate stages (\textit{hypotheses generation and data gathering for hypothesis testing}). These are the steps where physicians need more assistance. Our evaluation study results confirmed that the new system was able to improve the human-AI collaboration experience.

Clinicians (P2, P6) commented that the early prediction with uncertainty visualization and lab test recommendations could ease feelings of being in competition with or replaced by AI, since it no longer focuses on the final stage and leaves humans to decide whether to take the suggestions or not. Instead, our system ensures that physicians retain the role of final decision-maker.
\begin{quote}
    \textit{``if you tell me, you (model) need these labs, you got it, buddy, I will order those labs. It's not a problem...''} (P2)
\end{quote}

Our design probe significantly alleviated experts' concerns about the threat of AI. Participants felt that our new sepsis module offers the experience of teaming with AI rather than competition, creating an actual human-AI collaboration paradigm.
\begin{quote}
    \textit{``I think knowing what [lab results] would help AI to have a better prediction would let me give it a try, and [the lab results] may be more useful in my decision.''} (P6)
\end{quote}

The shift of AI assistance's focus and the improvement in the collaboration experience is combining efforts of multiple functions supported by our system. In the rest of this section, we summarize participants' feedback on each function of our system.

\subsubsection{Future Prediction and Visualization Reveals More Information}

All of the six participants in our study liked the time-based prediction capability of our system.
The prediction graph, with the uncertainty visualization, went beyond the final decision stage and provided much richer information than a risk score
\textit{``I love that graph because it says the whole picture.''} (P1)
This could help physicians better understand the model and inspect whether it is reliable. Participants agreed that adding more explanations to charts would help them make better decisions.
\begin{quote}
    \textit{``So I think having some of that [temporal] information is helpful to understand what drives the model and also where it's gonna go in 10 hours from now, if that model is good and consistent... I would hope that somebody could show that this where the model is gonna go [in the EPIC sepsis system], as it matches up with what happens in real life. ''} (P5)
\end{quote}

On the one hand, this is supported by previous work about the advantages and benefits of XAI \cite{wang2021explanations, zhang2020effect}. 
On the other hand, the future prediction function embeds the shift of AI focus from final diagnosis results to intermediate suggestions (help experts propose and test hypotheses). 
This leaves enough flexible space for human experts to make intermediate and final decisions.
Meanwhile, the system also places more emphasis on uncertainty estimation. 
The participants agreed that providing these aspects of information was much more helpful than a single risk score.
\begin{quote}
    \textit{``Yeah, I think that will be much better than just giving me just a score. If I can see something like this [counterfactual explanation], so the nurse gets this, or I am opening it up, I'm seeing that, okay, this patient came in with a risk score of five and now [with the suggested lab result] it is at 15. And based on the model, in the next eight hours, or 10 hours, it's going to be like 25 or 30. So, that is more meaningful information than just a score.'' } (P3)
\end{quote}

\subsubsection{Laboratory Test Recommendation Provides Actionable Insights}
All six participants responded positively to the lab test recommendation.
First of all, participants confirmed that the lab test recommendations were consistent with current physician workflows. 
This validates the design of this system function.
Meanwhile, participants unanimously said that they would take advantage of this recommendation function to help them consider and act on laboratory tests.
\begin{quote}
    \textit{``If I was able to click on the more lab tests needed and get an idea on how much it could narrow. And if it was something where it recommended these diagnostics, and I did these diagnostics, and when the labs came back, somehow triggered me to look at it again... and that prediction [uncertainty] has narrowed, [prediction] got very accurate in terms of risk score... This would be cool.''} (P2)
\end{quote}

Combined with the future prediction function, this could provide clinicians with more insights than a risk score or alert, helping them narrow down their hypotheses and make better sepsis diagnostic decisions.
\begin{quote}
    \textit{``In the ICU, that [lab recommendation] would be more realistic as a flag than the sepsis alert that we have right now, because it's telling you something. It's giving you information.''} (P1)
\end{quote}

These results suggest that our system design has the potential to provide actionable insights to health experts and support the stages from generating hypotheses to gathering data to testing hypotheses. 

\subsubsection{Counterfactual Information Improves Transparency}
Closely related to the laboratory test recommendation function, the participants agreed that showing counterfactual predictions based on recommendation would drive their choice of laboratory tests.
Traditionally, reliance on clinicians' personal experience was common, yet diagnosing sepsis posed significant challenges due to its inherent uncertainty.
With our system prototype, they commented that being able to show potential outcomes of different lab tests would help them with the thinking process and narrow down the search space.
\begin{quote}
    \textit{If the system is going to alert me: hey, maybe you should repeat a lactate now because your patient is scoring high; Or maybe you should repeat a white blood count today because you don't have lab work today; or something like that. That will certainly be helpful.} (P3)
\end{quote}

Moreover, the participants (P2, P4) also mentioned that showing counterfactual information also helped them better understand the model's functions and its future prediction process.

\begin{quote}
    \textit{``And then again, having it [the alternative trajectory with a counterfactual lab result] helps me with what AI wants me to order, so that it can be more accurate, [this design] is good.''} (P2)
\end{quote}

\begin{quote}
    \textit{``Especially if you're asking me to order a repeat lactate, telling me that, hey, your sepsis score was a five because your lactate was high and your white count was high. And I [AI] want that lactate, and [ordering a lactate] is going to help me kind of closing down my model. I think that'll be useful. ''} (P4)
\end{quote}


\subsubsection{Areas for Future Improvement}
\label{sub:stage2:future}
In addition to the positive comments through the design probe with our system, participants also suggested a few concerns and expectations about our system for future real-world deployment.
One concern is the performance of AI predictions and recommendations. With our algorithm having a stronger capability than a risk score, the model becomes more complex, and participants were concerned about its reliability. This is commonly observed in many health-focused AI systems~\cite{beede2020human,wang2021brilliant}.

Meanwhile, participant (P3) also mentioned the risk of information overload. \textit{``But simultaneously, you have to be careful and mindful of giving too much information at once. So maybe it can be a step-wise approach.''} (P3)
This suggests a future improvement direction of our system to that providing information step by step that follows clinicians through their decision-making process and provides appropriate explanations will best assist them in making a diagnosis.

In addition, some participants (P4, P5) also mentioned the potential to expand the usage scenarios of the counterfactual predictions and explanations, such as showing the potential impact of certain treatments on the development of a patient's symptom conditions. Although this is beyond the scope of this paper, we discuss a few promising directions of future work in Section~\ref{sec:discussion}.

\section{Discussion}
\label{sec:discussion}

In this section, we discuss the design implications obtained from our interview and evaluation studies. We then discuss the new human-AI collaboration paradigm and its application beyond sepsis diagnosis. We also highlight the risks and ethical concerns associated with the paradigm, as well as the limitations of our work.

\subsection{Human-AI Collaboration in High-Uncertainty, High-Stakes, and Time-Sensitive Decision Making}

With sepsis diagnosis as an example, our work reveals a fundamental problem in the existing human-AI collaboration paradigm for high-uncertainty, high-stakes, and time-sensitive decision-making tasks.
Most AI research in this space aims to create the most accurate risk score prediction model. 
However, this goal is unable to form an effective human-AI collaboration. 
Our study shows that physicians found such a risk prediction model unhelpful and intimidating, challenging their role as the final decision-maker.

Instead, we argue that AI should aim to support human experts in the intermediate stages (Step 1 - 3 in Figure~\ref{fig:human-AI-collab-paradigm}) rather than the final stage. 
This can position AI in an appropriate place to effectively support experts' decision-making process, while not influencing their decision-maker role.
In our case, we introduce a novel approach by providing future prediction and uncertainty visualization, lab test recommendations, and counterfactual information and support (rather than challenge) experts' final decisions.
This establishes a unique form of `communication and collaboration' between physicians and the AI by providing actionable recommendations.
Our evaluation study results suggest that such a shift in AI's focus can indeed create a new human-AI team paradigm. 
Our exploration into sepsis diagnostic decision-making exemplifies this collaborative process. 
In practice, this does not conflict with current XAI research and provides a new perspective on the role that AI can play in the decision-making process.

\subsection{Beyond Sepsis Diagnosis}
We envision such a new human-AI collaboration paradigm can move beyond sepsis diagnosis. 
In healthcare, there are a number of decision-making tasks that have similar properties as sepsis diagnosis: highly uncertain, high-stakes, and time-sensitive \cite{lai2021towards}.
Examples include both physical health problems (e.g., stroke, heart attack, and meningitis), and mental health problems (e.g., major depressive disorder with suicidal ideation, bipolar disorder during a manic episode, and schizophrenia with psychosis).
In these cases, the symptoms tend to be ambiguous, noisy, and highly individual, while having life-threatening consequences if not treated in a timely manner.
If an AI only focuses on predicting the outcome of the final decision stage, expert clinicians would also find it as a ``challenging and intimidating competitor'' rather than a collaborator and a partner, leading to the abortion of AI.
Our design can potentially be generalized to these fields, where AI should also support these experts in their intermediate decision-making stages and help them propose hypotheses, gather information, and test hypotheses.

In addition, we foresee that such a new human-AI collaboration paradigm can be applicable to other non-healthcare complex decision-making scenarios as well, such as military (e.g., hostage rescues, evacuation operations), business (e.g., product launch/recalls, market crisis), emergency response (e.g., earthquake response, wildfire response), just name a few.
All these cases require fast and accurate human decisions to reduce uncertainty and achieve optimal outcomes.
Our proposal of the new human-AI collaboration paradigm can inspire the existing solutions in these fields to shift their AI focus to better support domain experts.

\subsection{Risks and Ethical Concerns of AI-powered Decision Making}
Despite the promising advantage of our newly proposed human-AI collaboration paradigm, we also want to highlight the risks and ethical concerns associated with it.
For example, such a new paradigm would introduce additional burden to experts~\cite{mishra2021designing,10.1145/3290605.3300789,10.1145/3415224}.
In our study, participants were concerned about the cognitive load caused by our system, which has been reported in previous studies related to visualization \cite{jin2020carepre,wang2021brilliant, amershi2019guidelines}.
Meanwhile, mistakes and errors made by AI are inevitable, and the potential biases embedded in AI algorithms are not yet addressed by this new paradigm. 
The responsibility still falls on human experts to minimize these risks and biases through rigorous testing and evaluation. 
Besides, there is a potential risk of over-dependence on AI systems, which might foster complacency and reduce vigilance among human experts. 
This is an open research question for future researchers, requiring a balanced design approach that promotes collaboration while avoiding an undue reliance on AI recommendations.

\subsection{Limitation and Future Work}
There are several limitations in our work.
First, our study population is limited. 
We only involved six physicians in our formative study and heuristic evaluation (similar to the number of expert participants in prior work's formative study~\cite{cai2019hello}), who came from the same hospital and only used one specific sepsis module ESM. 
Although it is one of the mostly commonly used sepsis modules in the U.S., there could be some systematic biases in our study results. 
Future work needs to involve more diverse populations from multiple hospitals.
Second, our system is implemented as a prototype and not integrated into the EHR system. 
Our evaluation study used our system as a design probe to collect clinicians' feedback. 
This may influence the validity and generalizability of our results.
Our findings may be different if our system is actually deployed in the real world.
We picked ESM as a case study, since it has been the focus of extensive prior research as an early detection system \cite{cull2023epic}.
Further research is required to assess additional sepsis early detection systems and comparable tools for early disease identification to enhance understanding of the clinical decision-making process.
Moreover, to further elucidate the challenges arising from AI-assisted decision-making and to develop systems that are more congruent with the clinical decision-making processes, quantitative research and additional clinical practices are required in the future.
Third, our algorithm and visualization have room for improvement. 
As mentioned in Section~\ref{sec:system}, there are more visualization methods and algorithms for risk prediction and uncertainty estimation. 
Future work can explore the effectiveness of more back-end techniques.

Furthermore, in our interview, the participants mentioned that they barely relied on the existing sepsis module ESM in their decision-making process. Their comments reveal that this module needs a better design to support clinicians' workflow. Our prototype in Section~\ref{sec:system} presents an initial step towards a better design. And there are a few more directions to improve.
Based on the comments mentioned by the participants in Section~\ref{sub:stage2:future}, a future sepsis module should provide a simple and easy-to-operate interface. Clinicians are usually over-loaded~\cite{wang2021brilliant,jin2020carepre}, and an appropriately designed interface could improve their efficiency.

We also find that physicians and nurses often have different responsibilities in the diagnostic decision-making process.
In our case, nurses are tasked with receiving alerts and determining if they warrant escalation to physicians, while physicians make the ultimate decision on the necessary subsequent actions.
Different workflows that arise from different clinical roles are often overlooked in contemporary AI-CDSS designs.
However, predominant EHR systems and AI-CDSS platforms fail to distinguish between these distinct roles and tend to offer a one-size-fits-all interface and set of functionalities to both physicians and nurses.
In future designs, the systems should be role-specific and tailored to both physicians and nurses, ensuring that each can extract relevant information from model predictions. 
This not only enhances the system's efficiency but also ensures that each medical professional is equipped with the right tools to make informed decisions.

\section{Conclusion}

In this work, we aim to design a better human-AI collaboration paradigm to support human experts in high-uncertainty, high-stakes, and time-sensitive decision-making tasks.
We focus on sepsis diagnosis, a common yet life-threatening disease.
We conducted a formative study with six physicians with rich sepsis-treating experience to better understand the existing challenges of human-AI collaboration with a common sepsis module.
Our results reveal that the existing module is not only useless but also leads to additional meaningless workloads. 
More importantly, it reveals a fundamental problem of the wrong focus of AI: AI should not focus on predicting or suggesting the final decision-making stage, which could be challenging and intimidating to human experts as the final decision-maker.
Based on these insights, we developed a system that aims to address these challenges. 
Our new system, \system, shifts the AI's focus from the final stage to the intermediate stages (generating hypotheses, gathering information, and testing hypotheses). 
\system improves a sepsis diagnosis algorithm with future prediction and uncertainty visualization, provides lab test recommendations, and offers counterfactual information.
Our evaluation study shows that the new system prototype can provide actionable insights, improve transparency, and better support the clinicians' decision-making process, forming a new human-AI collaboration paradigm.
We envision that our findings can shed light on the design of better human-AI collaboration paradigms for other scenarios with complex decision-making tasks.

\begin{acks}
This work was funded in part by the National Science Foundation under award number IIS-2145625 and by the National Institutes of Health under award number R01GM141279 and R01MD018424. The content is solely the responsibility of the authors and does not necessarily represent the official views of the National Science Foundation or the National Institutes of Health.
The authors extend heartfelt thanks to the participants from OSU Wexner Medical Center.
\end{acks}
\clearpage
\bibliographystyle{ACM-Reference-Format}
\bibliography{sample-base}


\begin{thebibliography}{107}


\ifx \showCODEN    \undefined \def \showCODEN     #1{\unskip}     \fi
\ifx \showDOI      \undefined \def \showDOI       #1{#1}\fi
\ifx \showISBNx    \undefined \def \showISBNx     #1{\unskip}     \fi
\ifx \showISBNxiii \undefined \def \showISBNxiii  #1{\unskip}     \fi
\ifx \showISSN     \undefined \def \showISSN      #1{\unskip}     \fi
\ifx \showLCCN     \undefined \def \showLCCN      #1{\unskip}     \fi
\ifx \shownote     \undefined \def \shownote      #1{#1}          \fi
\ifx \showarticletitle \undefined \def \showarticletitle #1{#1}   \fi
\ifx \showURL      \undefined \def \showURL       {\relax}        \fi
\providecommand\bibfield[2]{#2}
\providecommand\bibinfo[2]{#2}
\providecommand\natexlab[1]{#1}
\providecommand\showeprint[2][]{arXiv:#2}

\bibitem[Amann et~al\mbox{.}(2022)]%
        {amann2022explain}
\bibfield{author}{\bibinfo{person}{Julia Amann}, \bibinfo{person}{Dennis Vetter}, \bibinfo{person}{Stig~Nikolaj Blomberg}, \bibinfo{person}{Helle~Collatz Christensen}, \bibinfo{person}{Megan Coffee}, \bibinfo{person}{Sara Gerke}, \bibinfo{person}{Thomas~K Gilbert}, \bibinfo{person}{Thilo Hagendorff}, \bibinfo{person}{Sune Holm}, \bibinfo{person}{Michelle Livne}, {et~al\mbox{.}}} \bibinfo{year}{2022}\natexlab{}.
\newblock \showarticletitle{To explain or not to explain?—Artificial intelligence explainability in clinical decision support systems}.
\newblock \bibinfo{journal}{\emph{PLOS Digital Health}} \bibinfo{volume}{1}, \bibinfo{number}{2} (\bibinfo{year}{2022}), \bibinfo{pages}{e0000016}.
\newblock


\bibitem[Amershi et~al\mbox{.}(2019)]%
        {amershi2019guidelines}
\bibfield{author}{\bibinfo{person}{Saleema Amershi}, \bibinfo{person}{Dan Weld}, \bibinfo{person}{Mihaela Vorvoreanu}, \bibinfo{person}{Adam Fourney}, \bibinfo{person}{Besmira Nushi}, \bibinfo{person}{Penny Collisson}, \bibinfo{person}{Jina Suh}, \bibinfo{person}{Shamsi Iqbal}, \bibinfo{person}{Paul~N. Bennett}, \bibinfo{person}{Kori Inkpen}, \bibinfo{person}{Jaime Teevan}, \bibinfo{person}{Ruth Kikin-Gil}, {and} \bibinfo{person}{Eric Horvitz}.} \bibinfo{year}{2019}\natexlab{}.
\newblock \showarticletitle{Guidelines for Human-AI Interaction}. In \bibinfo{booktitle}{\emph{Proceedings of the 2019 CHI Conference on Human Factors in Computing Systems}} (Glasgow, Scotland Uk) \emph{(\bibinfo{series}{CHI '19})}. \bibinfo{publisher}{Association for Computing Machinery}, \bibinfo{address}{New York, NY, USA}, \bibinfo{pages}{1–13}.
\newblock
\showISBNx{9781450359702}
\urldef\tempurl%
\url{https://doi.org/10.1145/3290605.3300233}
\showDOI{\tempurl}


\bibitem[Baytas et~al\mbox{.}(2017)]%
        {baytas2017patient}
\bibfield{author}{\bibinfo{person}{Inci~M Baytas}, \bibinfo{person}{Cao Xiao}, \bibinfo{person}{Xi Zhang}, \bibinfo{person}{Fei Wang}, \bibinfo{person}{Anil~K Jain}, {and} \bibinfo{person}{Jiayu Zhou}.} \bibinfo{year}{2017}\natexlab{}.
\newblock \showarticletitle{Patient subtyping via time-aware LSTM networks}. In \bibinfo{booktitle}{\emph{Proceedings of the 23rd ACM SIGKDD international conference on knowledge discovery and data mining}}. \bibinfo{pages}{65--74}.
\newblock


\bibitem[Bedoya et~al\mbox{.}(2019)]%
        {bedoya2019minimal}
\bibfield{author}{\bibinfo{person}{Armando~D Bedoya}, \bibinfo{person}{Meredith~E Clement}, \bibinfo{person}{Matthew Phelan}, \bibinfo{person}{Rebecca~C Steorts}, \bibinfo{person}{Cara O’Brien}, {and} \bibinfo{person}{Benjamin~A Goldstein}.} \bibinfo{year}{2019}\natexlab{}.
\newblock \showarticletitle{Minimal impact of implemented early warning score and best practice alert for patient deterioration}.
\newblock \bibinfo{journal}{\emph{Critical care medicine}} \bibinfo{volume}{47}, \bibinfo{number}{1} (\bibinfo{year}{2019}), \bibinfo{pages}{49}.
\newblock


\bibitem[Beede et~al\mbox{.}(2020)]%
        {beede2020human}
\bibfield{author}{\bibinfo{person}{Emma Beede}, \bibinfo{person}{Elizabeth Baylor}, \bibinfo{person}{Fred Hersch}, \bibinfo{person}{Anna Iurchenko}, \bibinfo{person}{Lauren Wilcox}, \bibinfo{person}{Paisan Ruamviboonsuk}, {and} \bibinfo{person}{Laura~M Vardoulakis}.} \bibinfo{year}{2020}\natexlab{}.
\newblock \showarticletitle{A human-centered evaluation of a deep learning system deployed in clinics for the detection of diabetic retinopathy}. In \bibinfo{booktitle}{\emph{Proceedings of the 2020 CHI conference on human factors in computing systems}}. \bibinfo{pages}{1--12}.
\newblock


\bibitem[Bone et~al\mbox{.}(1992)]%
        {bone1992definitions}
\bibfield{author}{\bibinfo{person}{Roger~C Bone}, \bibinfo{person}{Robert~A Balk}, \bibinfo{person}{Frank~B Cerra}, \bibinfo{person}{R~Phillip Dellinger}, \bibinfo{person}{Alan~M Fein}, \bibinfo{person}{William~A Knaus}, \bibinfo{person}{Roland~MH Schein}, {and} \bibinfo{person}{William~J Sibbald}.} \bibinfo{year}{1992}\natexlab{}.
\newblock \showarticletitle{Definitions for sepsis and organ failure and guidelines for the use of innovative therapies in sepsis}.
\newblock \bibinfo{journal}{\emph{Chest}} \bibinfo{volume}{101}, \bibinfo{number}{6} (\bibinfo{year}{1992}), \bibinfo{pages}{1644--1655}.
\newblock


\bibitem[Burgess et~al\mbox{.}(2023)]%
        {10.1145/3544548.3581251}
\bibfield{author}{\bibinfo{person}{Eleanor~R. Burgess}, \bibinfo{person}{Ivana Jankovic}, \bibinfo{person}{Melissa Austin}, \bibinfo{person}{Nancy Cai}, \bibinfo{person}{Adela Kapu\'{s}ci\'{n}ska}, \bibinfo{person}{Suzanne Currie}, \bibinfo{person}{J.~Marc Overhage}, \bibinfo{person}{Erika~S Poole}, {and} \bibinfo{person}{Jofish Kaye}.} \bibinfo{year}{2023}\natexlab{}.
\newblock \showarticletitle{Healthcare AI Treatment Decision Support: Design Principles to Enhance Clinician Adoption and Trust}. In \bibinfo{booktitle}{\emph{Proceedings of the 2023 CHI Conference on Human Factors in Computing Systems}} (Hamburg, Germany) \emph{(\bibinfo{series}{CHI '23})}. \bibinfo{publisher}{Association for Computing Machinery}, \bibinfo{address}{New York, NY, USA}, Article \bibinfo{articleno}{15}, \bibinfo{numpages}{19}~pages.
\newblock
\showISBNx{9781450394215}
\urldef\tempurl%
\url{https://doi.org/10.1145/3544548.3581251}
\showDOI{\tempurl}


\bibitem[Caballero-Ruiz et~al\mbox{.}(2017)]%
        {caballero2017web}
\bibfield{author}{\bibinfo{person}{Estefan{\'\i}a Caballero-Ruiz}, \bibinfo{person}{Gema Garc{\'\i}a-S{\'a}ez}, \bibinfo{person}{Mercedes Rigla}, \bibinfo{person}{Mar{\'\i}a Villaplana}, \bibinfo{person}{Belen Pons}, {and} \bibinfo{person}{M~Elena Hernando}.} \bibinfo{year}{2017}\natexlab{}.
\newblock \showarticletitle{A web-based clinical decision support system for gestational diabetes: Automatic diet prescription and detection of insulin needs}.
\newblock \bibinfo{journal}{\emph{International journal of medical informatics}}  \bibinfo{volume}{102} (\bibinfo{year}{2017}), \bibinfo{pages}{35--49}.
\newblock


\bibitem[Cai et~al\mbox{.}(2019a)]%
        {cai2019human}
\bibfield{author}{\bibinfo{person}{Carrie~J Cai}, \bibinfo{person}{Emily Reif}, \bibinfo{person}{Narayan Hegde}, \bibinfo{person}{Jason Hipp}, \bibinfo{person}{Been Kim}, \bibinfo{person}{Daniel Smilkov}, \bibinfo{person}{Martin Wattenberg}, \bibinfo{person}{Fernanda Viegas}, \bibinfo{person}{Greg~S Corrado}, \bibinfo{person}{Martin~C Stumpe}, {et~al\mbox{.}}} \bibinfo{year}{2019}\natexlab{a}.
\newblock \showarticletitle{Human-centered tools for coping with imperfect algorithms during medical decision-making}. In \bibinfo{booktitle}{\emph{Proceedings of the 2019 chi conference on human factors in computing systems}}. \bibinfo{pages}{1--14}.
\newblock


\bibitem[Cai et~al\mbox{.}(2019b)]%
        {cai2019hello}
\bibfield{author}{\bibinfo{person}{Carrie~J. Cai}, \bibinfo{person}{Samantha Winter}, \bibinfo{person}{David Steiner}, \bibinfo{person}{Lauren Wilcox}, {and} \bibinfo{person}{Michael Terry}.} \bibinfo{year}{2019}\natexlab{b}.
\newblock \showarticletitle{"Hello AI": Uncovering the Onboarding Needs of Medical Practitioners for Human-AI Collaborative Decision-Making}.
\newblock \bibinfo{journal}{\emph{Proc. ACM Hum.-Comput. Interact.}} \bibinfo{volume}{3}, \bibinfo{number}{CSCW}, Article \bibinfo{articleno}{104} (\bibinfo{date}{nov} \bibinfo{year}{2019}), \bibinfo{numpages}{24}~pages.
\newblock
\urldef\tempurl%
\url{https://doi.org/10.1145/3359206}
\showDOI{\tempurl}


\bibitem[Cecconi et~al\mbox{.}(2018)]%
        {cecconi2018sepsis}
\bibfield{author}{\bibinfo{person}{Maurizio Cecconi}, \bibinfo{person}{Laura Evans}, \bibinfo{person}{Mitchell Levy}, {and} \bibinfo{person}{Andrew Rhodes}.} \bibinfo{year}{2018}\natexlab{}.
\newblock \showarticletitle{Sepsis and septic shock}.
\newblock \bibinfo{journal}{\emph{The Lancet}} \bibinfo{volume}{392}, \bibinfo{number}{10141} (\bibinfo{year}{2018}), \bibinfo{pages}{75--87}.
\newblock


\bibitem[Chaddad et~al\mbox{.}(2023)]%
        {chaddad2023survey}
\bibfield{author}{\bibinfo{person}{Ahmad Chaddad}, \bibinfo{person}{Jihao Peng}, \bibinfo{person}{Jian Xu}, {and} \bibinfo{person}{Ahmed Bouridane}.} \bibinfo{year}{2023}\natexlab{}.
\newblock \showarticletitle{Survey of explainable AI techniques in healthcare}.
\newblock \bibinfo{journal}{\emph{Sensors}} \bibinfo{volume}{23}, \bibinfo{number}{2} (\bibinfo{year}{2023}), \bibinfo{pages}{634}.
\newblock


\bibitem[Cheng et~al\mbox{.}(2019)]%
        {10.1145/3290605.3300789}
\bibfield{author}{\bibinfo{person}{Hao-Fei Cheng}, \bibinfo{person}{Ruotong Wang}, \bibinfo{person}{Zheng Zhang}, \bibinfo{person}{Fiona O'Connell}, \bibinfo{person}{Terrance Gray}, \bibinfo{person}{F.~Maxwell Harper}, {and} \bibinfo{person}{Haiyi Zhu}.} \bibinfo{year}{2019}\natexlab{}.
\newblock \showarticletitle{Explaining Decision-Making Algorithms through UI: Strategies to Help Non-Expert Stakeholders}. In \bibinfo{booktitle}{\emph{Proceedings of the 2019 CHI Conference on Human Factors in Computing Systems}} (Glasgow, Scotland Uk) \emph{(\bibinfo{series}{CHI '19})}. \bibinfo{publisher}{Association for Computing Machinery}, \bibinfo{address}{New York, NY, USA}, \bibinfo{pages}{1–12}.
\newblock
\showISBNx{9781450359702}
\urldef\tempurl%
\url{https://doi.org/10.1145/3290605.3300789}
\showDOI{\tempurl}


\bibitem[Cheng et~al\mbox{.}(2016)]%
        {cheng2016risk}
\bibfield{author}{\bibinfo{person}{Yu Cheng}, \bibinfo{person}{Fei Wang}, \bibinfo{person}{Ping Zhang}, {and} \bibinfo{person}{Jianying Hu}.} \bibinfo{year}{2016}\natexlab{}.
\newblock \showarticletitle{Risk prediction with electronic health records: A deep learning approach}. In \bibinfo{booktitle}{\emph{Proceedings of the 2016 SIAM international conference on data mining}}. SIAM, \bibinfo{pages}{432--440}.
\newblock


\bibitem[Chiang et~al\mbox{.}(2023)]%
        {chiang2023two}
\bibfield{author}{\bibinfo{person}{Chun-Wei Chiang}, \bibinfo{person}{Zhuoran Lu}, \bibinfo{person}{Zhuoyan Li}, {and} \bibinfo{person}{Ming Yin}.} \bibinfo{year}{2023}\natexlab{}.
\newblock \showarticletitle{Are Two Heads Better Than One in AI-Assisted Decision Making? Comparing the Behavior and Performance of Groups and Individuals in Human-AI Collaborative Recidivism Risk Assessment}. In \bibinfo{booktitle}{\emph{Proceedings of the 2023 CHI Conference on Human Factors in Computing Systems}} (Hamburg, Germany) \emph{(\bibinfo{series}{CHI '23})}. \bibinfo{publisher}{Association for Computing Machinery}, \bibinfo{address}{New York, NY, USA}, Article \bibinfo{articleno}{348}, \bibinfo{numpages}{18}~pages.
\newblock
\showISBNx{9781450394215}
\urldef\tempurl%
\url{https://doi.org/10.1145/3544548.3581015}
\showDOI{\tempurl}


\bibitem[Choi et~al\mbox{.}(2017)]%
        {choi2017gram}
\bibfield{author}{\bibinfo{person}{Edward Choi}, \bibinfo{person}{Mohammad~Taha Bahadori}, \bibinfo{person}{Le Song}, \bibinfo{person}{Walter~F Stewart}, {and} \bibinfo{person}{Jimeng Sun}.} \bibinfo{year}{2017}\natexlab{}.
\newblock \showarticletitle{GRAM: graph-based attention model for healthcare representation learning}. In \bibinfo{booktitle}{\emph{Proceedings of the 23rd ACM SIGKDD international conference on knowledge discovery and data mining}}. \bibinfo{pages}{787--795}.
\newblock


\bibitem[Choi et~al\mbox{.}(2016)]%
        {choi2016retain}
\bibfield{author}{\bibinfo{person}{Edward Choi}, \bibinfo{person}{Mohammad~Taha Bahadori}, \bibinfo{person}{Jimeng Sun}, \bibinfo{person}{Joshua Kulas}, \bibinfo{person}{Andy Schuetz}, {and} \bibinfo{person}{Walter Stewart}.} \bibinfo{year}{2016}\natexlab{}.
\newblock \showarticletitle{Retain: An interpretable predictive model for healthcare using reverse time attention mechanism}.
\newblock \bibinfo{journal}{\emph{Advances in neural information processing systems}}  \bibinfo{volume}{29} (\bibinfo{year}{2016}).
\newblock


\bibitem[Cui et~al\mbox{.}(2020)]%
        {cui2020deterrent}
\bibfield{author}{\bibinfo{person}{Limeng Cui}, \bibinfo{person}{Haeseung Seo}, \bibinfo{person}{Maryam Tabar}, \bibinfo{person}{Fenglong Ma}, \bibinfo{person}{Suhang Wang}, {and} \bibinfo{person}{Dongwon Lee}.} \bibinfo{year}{2020}\natexlab{}.
\newblock \showarticletitle{DETERRENT: Knowledge Guided Graph Attention Network for Detecting Healthcare Misinformation}. In \bibinfo{booktitle}{\emph{Proceedings of the 26th ACM SIGKDD International Conference on Knowledge Discovery \& Data Mining}} (Virtual Event, CA, USA) \emph{(\bibinfo{series}{KDD '20})}. \bibinfo{publisher}{Association for Computing Machinery}, \bibinfo{address}{New York, NY, USA}, \bibinfo{pages}{492–502}.
\newblock
\showISBNx{9781450379984}
\urldef\tempurl%
\url{https://doi.org/10.1145/3394486.3403092}
\showDOI{\tempurl}


\bibitem[Cull et~al\mbox{.}(2023)]%
        {cull2023epic}
\bibfield{author}{\bibinfo{person}{John Cull}, \bibinfo{person}{Robert Brevetta}, \bibinfo{person}{Jeff Gerac}, \bibinfo{person}{Shanu Kothari}, {and} \bibinfo{person}{Dawn Blackhurst}.} \bibinfo{year}{2023}\natexlab{}.
\newblock \showarticletitle{Epic Sepsis Model Inpatient Predictive Analytic Tool: A Validation Study}.
\newblock \bibinfo{journal}{\emph{Critical Care Explorations}} \bibinfo{volume}{5}, \bibinfo{number}{7} (\bibinfo{year}{2023}).
\newblock


\bibitem[Dierig et~al\mbox{.}(2018)]%
        {dierig2018time}
\bibfield{author}{\bibinfo{person}{Alexa Dierig}, \bibinfo{person}{Christoph Berger}, \bibinfo{person}{Philipp K.~A. Agyeman}, \bibinfo{person}{Sara Bernhard-Stirnemann}, \bibinfo{person}{Eric Giannoni}, \bibinfo{person}{Martin Stocker}, \bibinfo{person}{Klara~M. Posfay-Barbe}, \bibinfo{person}{Anita Niederer-Loher}, \bibinfo{person}{Christian~R. Kahlert}, \bibinfo{person}{Alex Donas}, \bibinfo{person}{Paul Hasters}, \bibinfo{person}{Christa Relly}, \bibinfo{person}{Thomas Riedel}, \bibinfo{person}{Christoph Aebi}, \bibinfo{person}{Luregn~J. Schlapbach}, {and} \bibinfo{person}{Swiss Pediatric Sepsis~Study Heininger, Ulrich~and}.} \bibinfo{year}{2018}\natexlab{}.
\newblock \showarticletitle{Time-to-Positivity of Blood Cultures in Children With Sepsis}.
\newblock \bibinfo{journal}{\emph{Frontiers in Pediatrics}}  \bibinfo{volume}{6} (\bibinfo{year}{2018}).
\newblock
\showISSN{2296-2360}
\urldef\tempurl%
\url{https://doi.org/10.3389/fped.2018.00222}
\showDOI{\tempurl}


\bibitem[Dorsett et~al\mbox{.}(2017)]%
        {dorsett2017qsofa}
\bibfield{author}{\bibinfo{person}{Maia Dorsett}, \bibinfo{person}{Melissa Kroll}, \bibinfo{person}{Clark~S Smith}, \bibinfo{person}{Phillip Asaro}, \bibinfo{person}{Stephen~Y Liang}, {and} \bibinfo{person}{Hawnwan~P Moy}.} \bibinfo{year}{2017}\natexlab{}.
\newblock \showarticletitle{qSOFA has poor sensitivity for prehospital identification of severe sepsis and septic shock}.
\newblock \bibinfo{journal}{\emph{Prehospital emergency care}} \bibinfo{volume}{21}, \bibinfo{number}{4} (\bibinfo{year}{2017}), \bibinfo{pages}{489--497}.
\newblock


\bibitem[Downing et~al\mbox{.}(2019)]%
        {downing2019electronic}
\bibfield{author}{\bibinfo{person}{Norman~Lance Downing}, \bibinfo{person}{Joshua Rolnick}, \bibinfo{person}{Sarah~F Poole}, \bibinfo{person}{Evan Hall}, \bibinfo{person}{Alexander~J Wessels}, \bibinfo{person}{Paul Heidenreich}, {and} \bibinfo{person}{Lisa Shieh}.} \bibinfo{year}{2019}\natexlab{}.
\newblock \showarticletitle{Electronic health record-based clinical decision support alert for severe sepsis: a randomised evaluation}.
\newblock \bibinfo{journal}{\emph{BMJ quality \& safety}} \bibinfo{volume}{28}, \bibinfo{number}{9} (\bibinfo{year}{2019}), \bibinfo{pages}{762--768}.
\newblock


\bibitem[Du et~al\mbox{.}(2023)]%
        {9765710}
\bibfield{author}{\bibinfo{person}{Yongping Du}, \bibinfo{person}{Jingya Yan}, \bibinfo{person}{Yuxuan Lu}, \bibinfo{person}{Yiliang Zhao}, {and} \bibinfo{person}{Xingnan Jin}.} \bibinfo{year}{2023}\natexlab{}.
\newblock \showarticletitle{Improving Biomedical Question Answering by Data Augmentation and Model Weighting}.
\newblock \bibinfo{journal}{\emph{IEEE/ACM Transactions on Computational Biology and Bioinformatics}} \bibinfo{volume}{20}, \bibinfo{number}{2} (\bibinfo{year}{2023}), \bibinfo{pages}{1114--1124}.
\newblock
\urldef\tempurl%
\url{https://doi.org/10.1109/TCBB.2022.3171388}
\showDOI{\tempurl}


\bibitem[Dur{\'a}n and Jongsma(2021)]%
        {duran2021afraid}
\bibfield{author}{\bibinfo{person}{Juan~Manuel Dur{\'a}n} {and} \bibinfo{person}{Karin~Rolanda Jongsma}.} \bibinfo{year}{2021}\natexlab{}.
\newblock \showarticletitle{Who is afraid of black box algorithms? On the epistemological and ethical basis of trust in medical AI}.
\newblock \bibinfo{journal}{\emph{Journal of Medical Ethics}} \bibinfo{volume}{47}, \bibinfo{number}{5} (\bibinfo{year}{2021}), \bibinfo{pages}{329--335}.
\newblock
\showISSN{0306-6800}
\urldef\tempurl%
\url{https://doi.org/10.1136/medethics-2020-106820}
\showDOI{\tempurl}
\showeprint{https://jme.bmj.com/content/47/5/329.full.pdf}


\bibitem[Ehsan et~al\mbox{.}(2022)]%
        {10.1145/3491101.3503727}
\bibfield{author}{\bibinfo{person}{Upol Ehsan}, \bibinfo{person}{Philipp Wintersberger}, \bibinfo{person}{Q.~Vera Liao}, \bibinfo{person}{Elizabeth~Anne Watkins}, \bibinfo{person}{Carina Manger}, \bibinfo{person}{Hal Daum\'{e}~III}, \bibinfo{person}{Andreas Riener}, {and} \bibinfo{person}{Mark~O Riedl}.} \bibinfo{year}{2022}\natexlab{}.
\newblock \showarticletitle{Human-Centered Explainable AI (HCXAI): Beyond Opening the Black-Box of AI}. In \bibinfo{booktitle}{\emph{Extended Abstracts of the 2022 CHI Conference on Human Factors in Computing Systems}} (New Orleans, LA, USA) \emph{(\bibinfo{series}{CHI EA '22})}. \bibinfo{publisher}{Association for Computing Machinery}, \bibinfo{address}{New York, NY, USA}, Article \bibinfo{articleno}{109}, \bibinfo{numpages}{7}~pages.
\newblock
\showISBNx{9781450391566}
\urldef\tempurl%
\url{https://doi.org/10.1145/3491101.3503727}
\showDOI{\tempurl}


\bibitem[Evans et~al\mbox{.}(2021)]%
        {evans2021surviving}
\bibfield{author}{\bibinfo{person}{Laura Evans}, \bibinfo{person}{Andrew Rhodes}, \bibinfo{person}{Waleed Alhazzani}, \bibinfo{person}{Massimo Antonelli}, \bibinfo{person}{Craig~M Coopersmith}, \bibinfo{person}{Craig French}, \bibinfo{person}{Fl{\'a}via~R Machado}, \bibinfo{person}{Lauralyn Mcintyre}, \bibinfo{person}{Marlies Ostermann}, \bibinfo{person}{Hallie~C Prescott}, {et~al\mbox{.}}} \bibinfo{year}{2021}\natexlab{}.
\newblock \showarticletitle{Surviving sepsis campaign: international guidelines for management of sepsis and septic shock 2021}.
\newblock \bibinfo{journal}{\emph{Intensive care medicine}} \bibinfo{volume}{47}, \bibinfo{number}{11} (\bibinfo{year}{2021}), \bibinfo{pages}{1181--1247}.
\newblock


\bibitem[Fogliato et~al\mbox{.}(2022)]%
        {fogliato2022goes}
\bibfield{author}{\bibinfo{person}{Riccardo Fogliato}, \bibinfo{person}{Shreya Chappidi}, \bibinfo{person}{Matthew Lungren}, \bibinfo{person}{Paul Fisher}, \bibinfo{person}{Diane Wilson}, \bibinfo{person}{Michael Fitzke}, \bibinfo{person}{Mark Parkinson}, \bibinfo{person}{Eric Horvitz}, \bibinfo{person}{Kori Inkpen}, {and} \bibinfo{person}{Besmira Nushi}.} \bibinfo{year}{2022}\natexlab{}.
\newblock \showarticletitle{Who Goes First? Influences of Human-AI Workflow on Decision Making in Clinical Imaging}. In \bibinfo{booktitle}{\emph{Proceedings of the 2022 ACM Conference on Fairness, Accountability, and Transparency}} (Seoul, Republic of Korea) \emph{(\bibinfo{series}{FAccT '22})}. \bibinfo{publisher}{Association for Computing Machinery}, \bibinfo{address}{New York, NY, USA}, \bibinfo{pages}{1362–1374}.
\newblock
\showISBNx{9781450393522}
\urldef\tempurl%
\url{https://doi.org/10.1145/3531146.3533193}
\showDOI{\tempurl}


\bibitem[Ghassemi et~al\mbox{.}(2021)]%
        {ghassemi2021false}
\bibfield{author}{\bibinfo{person}{Marzyeh Ghassemi}, \bibinfo{person}{Luke Oakden-Rayner}, {and} \bibinfo{person}{Andrew~L Beam}.} \bibinfo{year}{2021}\natexlab{}.
\newblock \showarticletitle{The false hope of current approaches to explainable artificial intelligence in health care}.
\newblock \bibinfo{journal}{\emph{The Lancet Digital Health}} \bibinfo{volume}{3}, \bibinfo{number}{11} (\bibinfo{year}{2021}), \bibinfo{pages}{e745--e750}.
\newblock


\bibitem[Goh et~al\mbox{.}(2021)]%
        {goh2021artificial}
\bibfield{author}{\bibinfo{person}{Kim~Huat Goh}, \bibinfo{person}{Le Wang}, \bibinfo{person}{Adrian Yong~Kwang Yeow}, \bibinfo{person}{Hermione Poh}, \bibinfo{person}{Ke Li}, \bibinfo{person}{Joannas Jie~Lin Yeow}, {and} \bibinfo{person}{Gamaliel Yu~Heng Tan}.} \bibinfo{year}{2021}\natexlab{}.
\newblock \showarticletitle{Artificial intelligence in sepsis early prediction and diagnosis using unstructured data in healthcare}.
\newblock \bibinfo{journal}{\emph{Nature communications}} \bibinfo{volume}{12}, \bibinfo{number}{1} (\bibinfo{year}{2021}), \bibinfo{pages}{711}.
\newblock


\bibitem[Goodman(1961)]%
        {goodman1961snowball}
\bibfield{author}{\bibinfo{person}{Leo~A Goodman}.} \bibinfo{year}{1961}\natexlab{}.
\newblock \showarticletitle{Snowball sampling}.
\newblock \bibinfo{journal}{\emph{The annals of mathematical statistics}} (\bibinfo{year}{1961}), \bibinfo{pages}{148--170}.
\newblock


\bibitem[Gutierrez(2020)]%
        {gutierrez2020artificial}
\bibfield{author}{\bibinfo{person}{Guillermo Gutierrez}.} \bibinfo{year}{2020}\natexlab{}.
\newblock \showarticletitle{Artificial intelligence in the intensive care unit}.
\newblock \bibinfo{journal}{\emph{Annual Update in Intensive Care and Emergency Medicine 2020}} (\bibinfo{year}{2020}), \bibinfo{pages}{667--681}.
\newblock


\bibitem[Habib et~al\mbox{.}(2021)]%
        {10.1001/jamainternmed.2021.3333}
\bibfield{author}{\bibinfo{person}{Anand~R. Habib}, \bibinfo{person}{Anthony~L. Lin}, {and} \bibinfo{person}{Richard~W. Grant}.} \bibinfo{year}{2021}\natexlab{}.
\newblock \showarticletitle{{The Epic Sepsis Model Falls Short—The Importance of External Validation}}.
\newblock \bibinfo{journal}{\emph{JAMA Internal Medicine}} \bibinfo{volume}{181}, \bibinfo{number}{8} (\bibinfo{date}{08} \bibinfo{year}{2021}), \bibinfo{pages}{1040--1041}.
\newblock
\showISSN{2168-6106}
\urldef\tempurl%
\url{https://doi.org/10.1001/jamainternmed.2021.3333}
\showDOI{\tempurl}


\bibitem[Hochreiter and Schmidhuber(1997)]%
        {hochreiter1997long}
\bibfield{author}{\bibinfo{person}{Sepp Hochreiter} {and} \bibinfo{person}{J{\"u}rgen Schmidhuber}.} \bibinfo{year}{1997}\natexlab{}.
\newblock \showarticletitle{Long short-term memory}.
\newblock \bibinfo{journal}{\emph{Neural computation}} \bibinfo{volume}{9}, \bibinfo{number}{8} (\bibinfo{year}{1997}), \bibinfo{pages}{1735--1780}.
\newblock


\bibitem[Iftikhar et~al\mbox{.}(2020)]%
        {iftikhar2020artificial}
\bibfield{author}{\bibinfo{person}{Pulwasha Iftikhar}, \bibinfo{person}{Marcela~V Kuijpers}, \bibinfo{person}{Azadeh Khayyat}, \bibinfo{person}{Aqsa Iftikhar}, {and} \bibinfo{person}{Maribel~DeGouvia De~Sa}.} \bibinfo{year}{2020}\natexlab{}.
\newblock \showarticletitle{Artificial intelligence: a new paradigm in obstetrics and gynecology research and clinical practice}.
\newblock \bibinfo{journal}{\emph{Cureus}} \bibinfo{volume}{12}, \bibinfo{number}{2} (\bibinfo{year}{2020}).
\newblock


\bibitem[Jacobs et~al\mbox{.}(2021)]%
        {10.1145/3411764.3445385}
\bibfield{author}{\bibinfo{person}{Maia Jacobs}, \bibinfo{person}{Jeffrey He}, \bibinfo{person}{Melanie F.~Pradier}, \bibinfo{person}{Barbara Lam}, \bibinfo{person}{Andrew~C. Ahn}, \bibinfo{person}{Thomas~H. McCoy}, \bibinfo{person}{Roy~H. Perlis}, \bibinfo{person}{Finale Doshi-Velez}, {and} \bibinfo{person}{Krzysztof~Z. Gajos}.} \bibinfo{year}{2021}\natexlab{}.
\newblock \showarticletitle{Designing AI for Trust and Collaboration in Time-Constrained Medical Decisions: A Sociotechnical Lens}. In \bibinfo{booktitle}{\emph{Proceedings of the 2021 CHI Conference on Human Factors in Computing Systems}} (Yokohama, Japan) \emph{(\bibinfo{series}{CHI '21})}. \bibinfo{publisher}{Association for Computing Machinery}, \bibinfo{address}{New York, NY, USA}, Article \bibinfo{articleno}{659}, \bibinfo{numpages}{14}~pages.
\newblock
\showISBNx{9781450380966}
\urldef\tempurl%
\url{https://doi.org/10.1145/3411764.3445385}
\showDOI{\tempurl}


\bibitem[Jin et~al\mbox{.}(2020)]%
        {jin2020carepre}
\bibfield{author}{\bibinfo{person}{Zhuochen Jin}, \bibinfo{person}{Shuyuan Cui}, \bibinfo{person}{Shunan Guo}, \bibinfo{person}{David Gotz}, \bibinfo{person}{Jimeng Sun}, {and} \bibinfo{person}{Nan Cao}.} \bibinfo{year}{2020}\natexlab{}.
\newblock \showarticletitle{CarePre: An Intelligent Clinical Decision Assistance System}.
\newblock \bibinfo{journal}{\emph{ACM Trans. Comput. Healthcare}} \bibinfo{volume}{1}, \bibinfo{number}{1}, Article \bibinfo{articleno}{6} (\bibinfo{date}{mar} \bibinfo{year}{2020}), \bibinfo{numpages}{20}~pages.
\newblock
\urldef\tempurl%
\url{https://doi.org/10.1145/3344258}
\showDOI{\tempurl}


\bibitem[Johnson et~al\mbox{.}(2016)]%
        {johnson2016mimic}
\bibfield{author}{\bibinfo{person}{Alistair~EW Johnson}, \bibinfo{person}{Tom~J Pollard}, \bibinfo{person}{Lu Shen}, \bibinfo{person}{Li-wei~H Lehman}, \bibinfo{person}{Mengling Feng}, \bibinfo{person}{Mohammad Ghassemi}, \bibinfo{person}{Benjamin Moody}, \bibinfo{person}{Peter Szolovits}, \bibinfo{person}{Leo Anthony~Celi}, {and} \bibinfo{person}{Roger~G Mark}.} \bibinfo{year}{2016}\natexlab{}.
\newblock \showarticletitle{MIMIC-III, a freely accessible critical care database}.
\newblock \bibinfo{journal}{\emph{Scientific data}} \bibinfo{volume}{3}, \bibinfo{number}{1} (\bibinfo{year}{2016}), \bibinfo{pages}{1--9}.
\newblock


\bibitem[Juluru et~al\mbox{.}(2021)]%
        {juluru2021integrating}
\bibfield{author}{\bibinfo{person}{Krishna Juluru}, \bibinfo{person}{Hao-Hsin Shih}, \bibinfo{person}{Krishna~Nand Keshava~Murthy}, \bibinfo{person}{Pierre Elnajjar}, \bibinfo{person}{Amin El-Rowmeim}, \bibinfo{person}{Christopher Roth}, \bibinfo{person}{Brad Genereaux}, \bibinfo{person}{Josef Fox}, \bibinfo{person}{Eliot Siegel}, {and} \bibinfo{person}{Daniel~L Rubin}.} \bibinfo{year}{2021}\natexlab{}.
\newblock \showarticletitle{Integrating Al algorithms into the clinical workflow}.
\newblock \bibinfo{journal}{\emph{Radiology: Artificial Intelligence}} \bibinfo{volume}{3}, \bibinfo{number}{6} (\bibinfo{year}{2021}), \bibinfo{pages}{e210013}.
\newblock


\bibitem[Kamal et~al\mbox{.}(2020)]%
        {kamal2020interpretable}
\bibfield{author}{\bibinfo{person}{Sundreen~Asad Kamal}, \bibinfo{person}{Changchang Yin}, \bibinfo{person}{Buyue Qian}, {and} \bibinfo{person}{Ping Zhang}.} \bibinfo{year}{2020}\natexlab{}.
\newblock \showarticletitle{An interpretable risk prediction model for healthcare with pattern attention}.
\newblock \bibinfo{journal}{\emph{BMC Medical Informatics and Decision Making}}  \bibinfo{volume}{20} (\bibinfo{year}{2020}), \bibinfo{pages}{1--10}.
\newblock


\bibitem[Kim and Park(2019)]%
        {kim2019sepsis}
\bibfield{author}{\bibinfo{person}{Hwan~Il Kim} {and} \bibinfo{person}{Sunghoon Park}.} \bibinfo{year}{2019}\natexlab{}.
\newblock \showarticletitle{Sepsis: early recognition and optimized treatment}.
\newblock \bibinfo{journal}{\emph{Tuberculosis and respiratory diseases}} \bibinfo{volume}{82}, \bibinfo{number}{1} (\bibinfo{year}{2019}), \bibinfo{pages}{6--14}.
\newblock


\bibitem[Lai et~al\mbox{.}(2021)]%
        {lai2021towards}
\bibfield{author}{\bibinfo{person}{Vivian Lai}, \bibinfo{person}{Chacha Chen}, \bibinfo{person}{Q~Vera Liao}, \bibinfo{person}{Alison Smith-Renner}, {and} \bibinfo{person}{Chenhao Tan}.} \bibinfo{year}{2021}\natexlab{}.
\newblock \showarticletitle{Towards a science of human-ai decision making: a survey of empirical studies}.
\newblock \bibinfo{journal}{\emph{arXiv preprint arXiv:2112.11471}} (\bibinfo{year}{2021}).
\newblock


\bibitem[Lai and Tan(2019)]%
        {lai2019human}
\bibfield{author}{\bibinfo{person}{Vivian Lai} {and} \bibinfo{person}{Chenhao Tan}.} \bibinfo{year}{2019}\natexlab{}.
\newblock \showarticletitle{On Human Predictions with Explanations and Predictions of Machine Learning Models: A Case Study on Deception Detection}. In \bibinfo{booktitle}{\emph{Proceedings of the Conference on Fairness, Accountability, and Transparency}} (Atlanta, GA, USA) \emph{(\bibinfo{series}{FAT* '19})}. \bibinfo{publisher}{Association for Computing Machinery}, \bibinfo{address}{New York, NY, USA}, \bibinfo{pages}{29–38}.
\newblock
\showISBNx{9781450361255}
\urldef\tempurl%
\url{https://doi.org/10.1145/3287560.3287590}
\showDOI{\tempurl}


\bibitem[Lee et~al\mbox{.}(2021)]%
        {lee2021human}
\bibfield{author}{\bibinfo{person}{Min~Hun Lee}, \bibinfo{person}{Daniel~P. Siewiorek}, \bibinfo{person}{Asim Smailagic}, \bibinfo{person}{Alexandre Bernardino}, {and} \bibinfo{person}{Sergi~Berm\'{u}dez Berm\'{u}dez~i Badia}.} \bibinfo{year}{2021}\natexlab{}.
\newblock \showarticletitle{A Human-AI Collaborative Approach for Clinical Decision Making on Rehabilitation Assessment}. In \bibinfo{booktitle}{\emph{Proceedings of the 2021 CHI Conference on Human Factors in Computing Systems}} (Yokohama, Japan) \emph{(\bibinfo{series}{CHI '21})}. \bibinfo{publisher}{Association for Computing Machinery}, \bibinfo{address}{New York, NY, USA}, Article \bibinfo{articleno}{392}, \bibinfo{numpages}{14}~pages.
\newblock
\showISBNx{9781450380966}
\urldef\tempurl%
\url{https://doi.org/10.1145/3411764.3445472}
\showDOI{\tempurl}


\bibitem[Lee(2018)]%
        {lee2018understanding}
\bibfield{author}{\bibinfo{person}{Min~Kyung Lee}.} \bibinfo{year}{2018}\natexlab{}.
\newblock \showarticletitle{Understanding perception of algorithmic decisions: Fairness, trust, and emotion in response to algorithmic management}.
\newblock \bibinfo{journal}{\emph{Big Data \& Society}} \bibinfo{volume}{5}, \bibinfo{number}{1} (\bibinfo{year}{2018}), \bibinfo{pages}{2053951718756684}.
\newblock
\urldef\tempurl%
\url{https://doi.org/10.1177/2053951718756684}
\showDOI{\tempurl}
\showeprint{https://doi.org/10.1177/2053951718756684}


\bibitem[Lee and Rich(2021)]%
        {lee2021included}
\bibfield{author}{\bibinfo{person}{Min~Kyung Lee} {and} \bibinfo{person}{Katherine Rich}.} \bibinfo{year}{2021}\natexlab{}.
\newblock \showarticletitle{Who Is Included in Human Perceptions of AI?: Trust and Perceived Fairness around Healthcare AI and Cultural Mistrust}. In \bibinfo{booktitle}{\emph{Proceedings of the 2021 CHI Conference on Human Factors in Computing Systems}} (Yokohama, Japan) \emph{(\bibinfo{series}{CHI '21})}. \bibinfo{publisher}{Association for Computing Machinery}, \bibinfo{address}{New York, NY, USA}, Article \bibinfo{articleno}{138}, \bibinfo{numpages}{14}~pages.
\newblock
\showISBNx{9781450380966}
\urldef\tempurl%
\url{https://doi.org/10.1145/3411764.3445570}
\showDOI{\tempurl}


\bibitem[Lentzen et~al\mbox{.}(2022)]%
        {lentzen2022critical}
\bibfield{author}{\bibinfo{person}{Manuel Lentzen}, \bibinfo{person}{Sumit Madan}, \bibinfo{person}{Vanessa Lage-Rupprecht}, \bibinfo{person}{Lisa K{\"u}hnel}, \bibinfo{person}{Juliane Fluck}, \bibinfo{person}{Marc Jacobs}, \bibinfo{person}{Mirja Mittermaier}, \bibinfo{person}{Martin Witzenrath}, \bibinfo{person}{Peter Brunecker}, \bibinfo{person}{Martin Hofmann-Apitius}, {et~al\mbox{.}}} \bibinfo{year}{2022}\natexlab{}.
\newblock \showarticletitle{Critical assessment of transformer-based AI models for German clinical notes}.
\newblock \bibinfo{journal}{\emph{JAMIA open}} \bibinfo{volume}{5}, \bibinfo{number}{4} (\bibinfo{year}{2022}), \bibinfo{pages}{ooac087}.
\newblock


\bibitem[Liao et~al\mbox{.}(2020)]%
        {10.1145/3313831.3376590}
\bibfield{author}{\bibinfo{person}{Q.~Vera Liao}, \bibinfo{person}{Daniel Gruen}, {and} \bibinfo{person}{Sarah Miller}.} \bibinfo{year}{2020}\natexlab{}.
\newblock \showarticletitle{Questioning the AI: Informing Design Practices for Explainable AI User Experiences}. In \bibinfo{booktitle}{\emph{Proceedings of the 2020 CHI Conference on Human Factors in Computing Systems}} (Honolulu, HI, USA) \emph{(\bibinfo{series}{CHI '20})}. \bibinfo{publisher}{Association for Computing Machinery}, \bibinfo{address}{New York, NY, USA}, \bibinfo{pages}{1–15}.
\newblock
\showISBNx{9781450367080}
\urldef\tempurl%
\url{https://doi.org/10.1145/3313831.3376590}
\showDOI{\tempurl}


\bibitem[Lin et~al\mbox{.}(2018)]%
        {lin2018early}
\bibfield{author}{\bibinfo{person}{Chen Lin}, \bibinfo{person}{Yuan Zhang}, \bibinfo{person}{Julie Ivy}, \bibinfo{person}{Muge Capan}, \bibinfo{person}{Ryan Arnold}, \bibinfo{person}{Jeanne~M Huddleston}, {and} \bibinfo{person}{Min Chi}.} \bibinfo{year}{2018}\natexlab{}.
\newblock \showarticletitle{Early diagnosis and prediction of sepsis shock by combining static and dynamic information using convolutional-LSTM}. In \bibinfo{booktitle}{\emph{2018 IEEE international conference on healthcare informatics (ICHI)}}. IEEE, \bibinfo{pages}{219--228}.
\newblock


\bibitem[Liu et~al\mbox{.}(2019)]%
        {liu2019data}
\bibfield{author}{\bibinfo{person}{R. Liu}, \bibinfo{person}{J.L. Greenstein}, \bibinfo{person}{S.J. Granite}, \bibinfo{person}{J.C. Fackler}, \bibinfo{person}{M.M. Bembea}, \bibinfo{person}{S.V. Sarma}, {and} \bibinfo{person}{R.L. Winslow}.} \bibinfo{year}{2019}\natexlab{}.
\newblock \showarticletitle{Data-driven discovery of a novel sepsis pre-shock state predicts impending septic shock in the {ICU}}.
\newblock \bibinfo{journal}{\emph{Sci.~Rep.}} \bibinfo{volume}{9}, \bibinfo{number}{1} (\bibinfo{year}{2019}), \bibinfo{pages}{1--9}.
\newblock
\urldef\tempurl%
\url{https://doi.org/10.1038/s41598-019-42637-5}
\showDOI{\tempurl}


\bibitem[Longhurst(2003)]%
        {longhurst2003semi}
\bibfield{author}{\bibinfo{person}{Robyn Longhurst}.} \bibinfo{year}{2003}\natexlab{}.
\newblock \showarticletitle{Semi-structured interviews and focus groups}.
\newblock \bibinfo{journal}{\emph{Key methods in geography}} \bibinfo{volume}{3}, \bibinfo{number}{2} (\bibinfo{year}{2003}), \bibinfo{pages}{143--156}.
\newblock


\bibitem[Lu et~al\mbox{.}(2022)]%
        {lu2022contextual}
\bibfield{author}{\bibinfo{person}{Yuxuan Lu}, \bibinfo{person}{Jingya Yan}, \bibinfo{person}{Zhixuan Qi}, \bibinfo{person}{Zhongzheng Ge}, {and} \bibinfo{person}{Yongping Du}.} \bibinfo{year}{2022}\natexlab{}.
\newblock \showarticletitle{Contextual Embedding and Model Weighting by Fusing Domain Knowledge on Biomedical Question Answering}. In \bibinfo{booktitle}{\emph{Proceedings of the 13th ACM International Conference on Bioinformatics, Computational Biology and Health Informatics}} (Northbrook, Illinois) \emph{(\bibinfo{series}{BCB '22})}. \bibinfo{publisher}{Association for Computing Machinery}, \bibinfo{address}{New York, NY, USA}, Article \bibinfo{articleno}{54}, \bibinfo{numpages}{4}~pages.
\newblock
\showISBNx{9781450393867}
\urldef\tempurl%
\url{https://doi.org/10.1145/3535508.3545508}
\showDOI{\tempurl}


\bibitem[Lu and Yin(2021)]%
        {lu2021human}
\bibfield{author}{\bibinfo{person}{Zhuoran Lu} {and} \bibinfo{person}{Ming Yin}.} \bibinfo{year}{2021}\natexlab{}.
\newblock \showarticletitle{Human Reliance on Machine Learning Models When Performance Feedback is Limited: Heuristics and Risks}. In \bibinfo{booktitle}{\emph{Proceedings of the 2021 CHI Conference on Human Factors in Computing Systems}} (Yokohama, Japan) \emph{(\bibinfo{series}{CHI '21})}. \bibinfo{publisher}{Association for Computing Machinery}, \bibinfo{address}{New York, NY, USA}, Article \bibinfo{articleno}{78}, \bibinfo{numpages}{16}~pages.
\newblock
\showISBNx{9781450380966}
\urldef\tempurl%
\url{https://doi.org/10.1145/3411764.3445562}
\showDOI{\tempurl}


\bibitem[Lyons et~al\mbox{.}(2023)]%
        {lyons2023factors}
\bibfield{author}{\bibinfo{person}{Patrick~G Lyons}, \bibinfo{person}{Mackenzie~R Hofford}, \bibinfo{person}{C~Yu Sean}, \bibinfo{person}{Andrew~P Michelson}, \bibinfo{person}{Philip~RO Payne}, \bibinfo{person}{Catherine~L Hough}, {and} \bibinfo{person}{Karandeep Singh}.} \bibinfo{year}{2023}\natexlab{}.
\newblock \showarticletitle{Factors associated with variability in the performance of a proprietary sepsis prediction model across 9 networked hospitals in the US}.
\newblock \bibinfo{journal}{\emph{JAMA Internal Medicine}} (\bibinfo{year}{2023}).
\newblock


\bibitem[Ma et~al\mbox{.}(2017)]%
        {ma2017dipole}
\bibfield{author}{\bibinfo{person}{Fenglong Ma}, \bibinfo{person}{Radha Chitta}, \bibinfo{person}{Jing Zhou}, \bibinfo{person}{Quanzeng You}, \bibinfo{person}{Tong Sun}, {and} \bibinfo{person}{Jing Gao}.} \bibinfo{year}{2017}\natexlab{}.
\newblock \showarticletitle{Dipole: Diagnosis prediction in healthcare via attention-based bidirectional recurrent neural networks}. In \bibinfo{booktitle}{\emph{Proceedings of the 23rd ACM SIGKDD international conference on knowledge discovery and data mining}}. \bibinfo{pages}{1903--1911}.
\newblock


\bibitem[Ma et~al\mbox{.}(2018)]%
        {ma2018kame}
\bibfield{author}{\bibinfo{person}{Fenglong Ma}, \bibinfo{person}{Quanzeng You}, \bibinfo{person}{Houping Xiao}, \bibinfo{person}{Radha Chitta}, \bibinfo{person}{Jing Zhou}, {and} \bibinfo{person}{Jing Gao}.} \bibinfo{year}{2018}\natexlab{}.
\newblock \showarticletitle{Kame: Knowledge-based attention model for diagnosis prediction in healthcare}. In \bibinfo{booktitle}{\emph{Proceedings of the 27th ACM International Conference on Information and Knowledge Management}}. \bibinfo{pages}{743--752}.
\newblock


\bibitem[Mishra and Rzeszotarski(2021)]%
        {mishra2021designing}
\bibfield{author}{\bibinfo{person}{Swati Mishra} {and} \bibinfo{person}{Jeffrey~M Rzeszotarski}.} \bibinfo{year}{2021}\natexlab{}.
\newblock \showarticletitle{Designing Interactive Transfer Learning Tools for ML Non-Experts}. In \bibinfo{booktitle}{\emph{Proceedings of the 2021 CHI Conference on Human Factors in Computing Systems}} (Yokohama, Japan) \emph{(\bibinfo{series}{CHI '21})}. \bibinfo{publisher}{Association for Computing Machinery}, \bibinfo{address}{New York, NY, USA}, Article \bibinfo{articleno}{364}, \bibinfo{numpages}{15}~pages.
\newblock
\showISBNx{9781450380966}
\urldef\tempurl%
\url{https://doi.org/10.1145/3411764.3445096}
\showDOI{\tempurl}


\bibitem[Nguyen et~al\mbox{.}(2014)]%
        {nguyen2014automated}
\bibfield{author}{\bibinfo{person}{Su~Q Nguyen}, \bibinfo{person}{Edwin Mwakalindile}, \bibinfo{person}{James~S Booth}, \bibinfo{person}{Vicki Hogan}, \bibinfo{person}{Jordan Morgan}, \bibinfo{person}{Charles~T Prickett}, \bibinfo{person}{John~P Donnelly}, {and} \bibinfo{person}{Henry~E Wang}.} \bibinfo{year}{2014}\natexlab{}.
\newblock \showarticletitle{Automated electronic medical record sepsis detection in the emergency department}.
\newblock \bibinfo{journal}{\emph{PeerJ}}  \bibinfo{volume}{2} (\bibinfo{year}{2014}), \bibinfo{pages}{e343}.
\newblock


\bibitem[Otaki et~al\mbox{.}(2022)]%
        {otaki2022clinical}
\bibfield{author}{\bibinfo{person}{Yuka Otaki}, \bibinfo{person}{Ananya Singh}, \bibinfo{person}{Paul Kavanagh}, \bibinfo{person}{Robert~JH Miller}, \bibinfo{person}{Tejas Parekh}, \bibinfo{person}{Balaji~K Tamarappoo}, \bibinfo{person}{Tali Sharir}, \bibinfo{person}{Andrew~J Einstein}, \bibinfo{person}{Mathews~B Fish}, \bibinfo{person}{Terrence~D Ruddy}, {et~al\mbox{.}}} \bibinfo{year}{2022}\natexlab{}.
\newblock \showarticletitle{Clinical deployment of explainable artificial intelligence of SPECT for diagnosis of coronary artery disease}.
\newblock \bibinfo{journal}{\emph{Cardiovascular Imaging}} \bibinfo{volume}{15}, \bibinfo{number}{6} (\bibinfo{year}{2022}), \bibinfo{pages}{1091--1102}.
\newblock


\bibitem[Padilla et~al\mbox{.}(2022)]%
        {padilla2022}
\bibfield{author}{\bibinfo{person}{Lace Padilla}, \bibinfo{person}{Racquel Fygenson}, \bibinfo{person}{Spencer~C Castro}, {and} \bibinfo{person}{Enrico Bertini}.} \bibinfo{year}{2022}\natexlab{}.
\newblock \showarticletitle{Multiple forecast visualizations (mfvs): Trade-offs in trust and performance in multiple covid-19 forecast visualizations}.
\newblock \bibinfo{journal}{\emph{IEEE Transactions on Visualization and Computer Graphics}} \bibinfo{volume}{29}, \bibinfo{number}{1} (\bibinfo{year}{2022}), \bibinfo{pages}{12--22}.
\newblock


\bibitem[Padilla et~al\mbox{.}(2021)]%
        {padilla2021}
\bibfield{author}{\bibinfo{person}{Lace Padilla}, \bibinfo{person}{Matthew Kay}, {and} \bibinfo{person}{Jessica Hullman}.} \bibinfo{year}{2021}\natexlab{}.
\newblock \bibinfo{booktitle}{\emph{Uncertainty Visualization}}.
\newblock \bibinfo{publisher}{John Wiley \& Sons, Ltd}, \bibinfo{pages}{1--18}.
\newblock
\showISBNx{9781118445112}
\urldef\tempurl%
\url{https://doi.org/10.1002/9781118445112.stat08296}
\showDOI{\tempurl}
\showeprint{https://onlinelibrary.wiley.com/doi/pdf/10.1002/9781118445112.stat08296}


\bibitem[Paszke et~al\mbox{.}(2019)]%
        {paszke2019pytorch}
\bibfield{author}{\bibinfo{person}{Adam Paszke}, \bibinfo{person}{Sam Gross}, \bibinfo{person}{Francisco Massa}, \bibinfo{person}{Adam Lerer}, \bibinfo{person}{James Bradbury}, \bibinfo{person}{Gregory Chanan}, \bibinfo{person}{Trevor Killeen}, \bibinfo{person}{Zeming Lin}, \bibinfo{person}{Natalia Gimelshein}, \bibinfo{person}{Luca Antiga}, \bibinfo{person}{Alban Desmaison}, \bibinfo{person}{Andreas Kopf}, \bibinfo{person}{Edward Yang}, \bibinfo{person}{Zachary DeVito}, \bibinfo{person}{Martin Raison}, \bibinfo{person}{Alykhan Tejani}, \bibinfo{person}{Sasank Chilamkurthy}, \bibinfo{person}{Benoit Steiner}, \bibinfo{person}{Lu Fang}, \bibinfo{person}{Junjie Bai}, {and} \bibinfo{person}{Soumith Chintala}.} \bibinfo{year}{2019}\natexlab{}.
\newblock \showarticletitle{PyTorch: An Imperative Style, High-Performance Deep Learning Library}. In \bibinfo{booktitle}{\emph{Advances in Neural Information Processing Systems}}, \bibfield{editor}{\bibinfo{person}{H.~Wallach}, \bibinfo{person}{H.~Larochelle}, \bibinfo{person}{A.~Beygelzimer}, \bibinfo{person}{F.~d\textquotesingle Alch\'{e}-Buc}, \bibinfo{person}{E.~Fox}, {and} \bibinfo{person}{R.~Garnett}} (Eds.), Vol.~\bibinfo{volume}{32}. \bibinfo{publisher}{Curran Associates, Inc.}
\newblock


\bibitem[Price(2018)]%
        {price2018big}
\bibfield{author}{\bibinfo{person}{W~Nicholson Price}.} \bibinfo{year}{2018}\natexlab{}.
\newblock \showarticletitle{Big data and black-box medical algorithms}.
\newblock \bibinfo{journal}{\emph{Science translational medicine}} \bibinfo{volume}{10}, \bibinfo{number}{471} (\bibinfo{year}{2018}), \bibinfo{pages}{eaao5333}.
\newblock


\bibitem[Rafiei et~al\mbox{.}(2021)]%
        {rafiei2021ssp}
\bibfield{author}{\bibinfo{person}{Alireza Rafiei}, \bibinfo{person}{Alireza Rezaee}, \bibinfo{person}{Farshid Hajati}, \bibinfo{person}{Soheila Gheisari}, {and} \bibinfo{person}{Mojtaba Golzan}.} \bibinfo{year}{2021}\natexlab{}.
\newblock \showarticletitle{SSP: Early prediction of sepsis using fully connected LSTM-CNN model}.
\newblock \bibinfo{journal}{\emph{Computers in biology and medicine}}  \bibinfo{volume}{128} (\bibinfo{year}{2021}), \bibinfo{pages}{104110}.
\newblock


\bibitem[Rajkomar et~al\mbox{.}(2019)]%
        {rajkomar2019machine}
\bibfield{author}{\bibinfo{person}{Alvin Rajkomar}, \bibinfo{person}{Jeffrey Dean}, {and} \bibinfo{person}{Isaac Kohane}.} \bibinfo{year}{2019}\natexlab{}.
\newblock \showarticletitle{Machine learning in medicine}.
\newblock \bibinfo{journal}{\emph{New England Journal of Medicine}} \bibinfo{volume}{380}, \bibinfo{number}{14} (\bibinfo{year}{2019}), \bibinfo{pages}{1347--1358}.
\newblock


\bibitem[Romero-Brufau et~al\mbox{.}(2020)]%
        {romero2020lesson}
\bibfield{author}{\bibinfo{person}{Santiago Romero-Brufau}, \bibinfo{person}{Kirk~D Wyatt}, \bibinfo{person}{Patricia Boyum}, \bibinfo{person}{Mindy Mickelson}, \bibinfo{person}{Matthew Moore}, {and} \bibinfo{person}{Cheristi Cognetta-Rieke}.} \bibinfo{year}{2020}\natexlab{}.
\newblock \showarticletitle{A lesson in implementation: a pre-post study of providers’ experience with artificial intelligence-based clinical decision support}.
\newblock \bibinfo{journal}{\emph{International journal of medical informatics}}  \bibinfo{volume}{137} (\bibinfo{year}{2020}), \bibinfo{pages}{104072}.
\newblock


\bibitem[Rudd et~al\mbox{.}(2020)]%
        {rudd2020global}
\bibfield{author}{\bibinfo{person}{Kristina~E Rudd}, \bibinfo{person}{Sarah~Charlotte Johnson}, \bibinfo{person}{Kareha~M Agesa}, \bibinfo{person}{Katya~Anne Shackelford}, \bibinfo{person}{Derrick Tsoi}, \bibinfo{person}{Daniel~Rhodes Kievlan}, \bibinfo{person}{Danny~V Colombara}, \bibinfo{person}{Kevin~S Ikuta}, \bibinfo{person}{Niranjan Kissoon}, \bibinfo{person}{Simon Finfer}, {et~al\mbox{.}}} \bibinfo{year}{2020}\natexlab{}.
\newblock \showarticletitle{Global, regional, and national sepsis incidence and mortality, 1990--2017: analysis for the Global Burden of Disease Study}.
\newblock \bibinfo{journal}{\emph{The Lancet}} \bibinfo{volume}{395}, \bibinfo{number}{10219} (\bibinfo{year}{2020}), \bibinfo{pages}{200--211}.
\newblock


\bibitem[Saqib et~al\mbox{.}(2018)]%
        {saqib2018early}
\bibfield{author}{\bibinfo{person}{Mohammed Saqib}, \bibinfo{person}{Ying Sha}, {and} \bibinfo{person}{May~D Wang}.} \bibinfo{year}{2018}\natexlab{}.
\newblock \showarticletitle{Early prediction of sepsis in EMR records using traditional ML techniques and deep learning LSTM networks}. In \bibinfo{booktitle}{\emph{2018 40th Annual International Conference of the IEEE Engineering in Medicine and Biology Society (EMBC)}}. IEEE, \bibinfo{pages}{4038--4041}.
\newblock


\bibitem[Sendak et~al\mbox{.}(2020a)]%
        {sendak2020human}
\bibfield{author}{\bibinfo{person}{Mark Sendak}, \bibinfo{person}{Madeleine~Clare Elish}, \bibinfo{person}{Michael Gao}, \bibinfo{person}{Joseph Futoma}, \bibinfo{person}{William Ratliff}, \bibinfo{person}{Marshall Nichols}, \bibinfo{person}{Armando Bedoya}, \bibinfo{person}{Suresh Balu}, {and} \bibinfo{person}{Cara O'Brien}.} \bibinfo{year}{2020}\natexlab{a}.
\newblock \showarticletitle{" The human body is a black box" supporting clinical decision-making with deep learning}. In \bibinfo{booktitle}{\emph{Proceedings of the 2020 conference on fairness, accountability, and transparency}}. \bibinfo{pages}{99--109}.
\newblock


\bibitem[Sendak et~al\mbox{.}(2020b)]%
        {10.1145/3351095.3372827}
\bibfield{author}{\bibinfo{person}{Mark Sendak}, \bibinfo{person}{Madeleine~Clare Elish}, \bibinfo{person}{Michael Gao}, \bibinfo{person}{Joseph Futoma}, \bibinfo{person}{William Ratliff}, \bibinfo{person}{Marshall Nichols}, \bibinfo{person}{Armando Bedoya}, \bibinfo{person}{Suresh Balu}, {and} \bibinfo{person}{Cara O'Brien}.} \bibinfo{year}{2020}\natexlab{b}.
\newblock \showarticletitle{"The Human Body is a Black Box": Supporting Clinical Decision-Making with Deep Learning}. In \bibinfo{booktitle}{\emph{Proceedings of the 2020 Conference on Fairness, Accountability, and Transparency}} (Barcelona, Spain) \emph{(\bibinfo{series}{FAT* '20})}. \bibinfo{publisher}{Association for Computing Machinery}, \bibinfo{address}{New York, NY, USA}, \bibinfo{pages}{99–109}.
\newblock
\showISBNx{9781450369367}
\urldef\tempurl%
\url{https://doi.org/10.1145/3351095.3372827}
\showDOI{\tempurl}


\bibitem[Shah and Lee(2019)]%
        {shah2019ai}
\bibfield{author}{\bibinfo{person}{Nirav~R Shah} {and} \bibinfo{person}{Thomas~H Lee}.} \bibinfo{year}{2019}\natexlab{}.
\newblock \showarticletitle{What AI means for doctors and doctoring}.
\newblock \bibinfo{journal}{\emph{NEJM Catalyst}} \bibinfo{volume}{5}, \bibinfo{number}{5} (\bibinfo{year}{2019}).
\newblock


\bibitem[Shannon and Weaver(1949)]%
        {shannon1949mathematical}
\bibfield{author}{\bibinfo{person}{Claude~E Shannon} {and} \bibinfo{person}{Warren Weaver}.} \bibinfo{year}{1949}\natexlab{}.
\newblock \showarticletitle{A mathematical model of communication}.
\newblock \bibinfo{journal}{\emph{Urbana, IL: University of Illinois Press}}  \bibinfo{volume}{11} (\bibinfo{year}{1949}), \bibinfo{pages}{11--20}.
\newblock


\bibitem[Shen et~al\mbox{.}(2020)]%
        {10.1145/3415224}
\bibfield{author}{\bibinfo{person}{Hong Shen}, \bibinfo{person}{Haojian Jin}, \bibinfo{person}{\'{A}ngel~Alexander Cabrera}, \bibinfo{person}{Adam Perer}, \bibinfo{person}{Haiyi Zhu}, {and} \bibinfo{person}{Jason~I. Hong}.} \bibinfo{year}{2020}\natexlab{}.
\newblock \showarticletitle{Designing Alternative Representations of Confusion Matrices to Support Non-Expert Public Understanding of Algorithm Performance}.
\newblock \bibinfo{journal}{\emph{Proc. ACM Hum.-Comput. Interact.}} \bibinfo{volume}{4}, \bibinfo{number}{CSCW2}, Article \bibinfo{articleno}{153} (\bibinfo{date}{oct} \bibinfo{year}{2020}), \bibinfo{numpages}{22}~pages.
\newblock
\urldef\tempurl%
\url{https://doi.org/10.1145/3415224}
\showDOI{\tempurl}


\bibitem[Shneiderman and Plaisant(2010)]%
        {shneiderman2010designing}
\bibfield{author}{\bibinfo{person}{Ben Shneiderman} {and} \bibinfo{person}{Catherine Plaisant}.} \bibinfo{year}{2010}\natexlab{}.
\newblock \bibinfo{booktitle}{\emph{Designing the user interface: Strategies for effective human-computer interaction}}.
\newblock \bibinfo{publisher}{Pearson Education India}.
\newblock


\bibitem[Silvestri et~al\mbox{.}(2022)]%
        {silvestri2022desired}
\bibfield{author}{\bibinfo{person}{Jasmine~A Silvestri}, \bibinfo{person}{Tyler~E Kmiec}, \bibinfo{person}{Nicholas~S Bishop}, \bibinfo{person}{Susan~H Regli}, {and} \bibinfo{person}{Gary~E Weissman}.} \bibinfo{year}{2022}\natexlab{}.
\newblock \showarticletitle{Desired Characteristics of a Clinical Decision Support System for Early Sepsis Recognition: Interview Study Among Hospital-Based Clinicians}.
\newblock \bibinfo{journal}{\emph{JMIR Hum Factors}} \bibinfo{volume}{9}, \bibinfo{number}{4} (\bibinfo{date}{21 Oct} \bibinfo{year}{2022}), \bibinfo{pages}{e36976}.
\newblock
\showISSN{2292-9495}
\urldef\tempurl%
\url{https://doi.org/10.2196/36976}
\showDOI{\tempurl}


\bibitem[Singer et~al\mbox{.}(2016)]%
        {singer2016third}
\bibfield{author}{\bibinfo{person}{Mervyn Singer}, \bibinfo{person}{Clifford~S. Deutschman}, \bibinfo{person}{Christopher~Warren Seymour}, \bibinfo{person}{Manu Shankar-Hari}, \bibinfo{person}{Djillali Annane}, \bibinfo{person}{Michael Bauer}, \bibinfo{person}{Rinaldo Bellomo}, \bibinfo{person}{Gordon~R. Bernard}, \bibinfo{person}{Jean-Daniel Chiche}, \bibinfo{person}{Craig~M. Coopersmith}, \bibinfo{person}{Richard~S. Hotchkiss}, \bibinfo{person}{Mitchell~M. Levy}, \bibinfo{person}{John~C. Marshall}, \bibinfo{person}{Greg~S. Martin}, \bibinfo{person}{Steven~M. Opal}, \bibinfo{person}{Gordon~D. Rubenfeld}, \bibinfo{person}{Tom van~der Poll}, \bibinfo{person}{Jean-Louis Vincent}, {and} \bibinfo{person}{Derek~C. Angus}.} \bibinfo{year}{2016}\natexlab{}.
\newblock \showarticletitle{{The Third International Consensus Definitions for Sepsis and Septic Shock (Sepsis-3)}}.
\newblock \bibinfo{journal}{\emph{JAMA}} \bibinfo{volume}{315}, \bibinfo{number}{8} (\bibinfo{date}{02} \bibinfo{year}{2016}), \bibinfo{pages}{801--810}.
\newblock
\showISSN{0098-7484}
\urldef\tempurl%
\url{https://doi.org/10.1001/jama.2016.0287}
\showDOI{\tempurl}
\showeprint{https://jamanetwork.com/journals/jama/articlepdf/2492881/jsc160002.pdf}


\bibitem[Sivaraman et~al\mbox{.}(2023)]%
        {sivaraman2023ignore}
\bibfield{author}{\bibinfo{person}{Venkatesh Sivaraman}, \bibinfo{person}{Leigh~A Bukowski}, \bibinfo{person}{Joel Levin}, \bibinfo{person}{Jeremy~M. Kahn}, {and} \bibinfo{person}{Adam Perer}.} \bibinfo{year}{2023}\natexlab{}.
\newblock \showarticletitle{Ignore, Trust, or Negotiate: Understanding Clinician Acceptance of AI-Based Treatment Recommendations in Health Care}. In \bibinfo{booktitle}{\emph{Proceedings of the 2023 CHI Conference on Human Factors in Computing Systems}} (Hamburg, Germany) \emph{(\bibinfo{series}{CHI '23})}. \bibinfo{publisher}{Association for Computing Machinery}, \bibinfo{address}{New York, NY, USA}, Article \bibinfo{articleno}{754}, \bibinfo{numpages}{18}~pages.
\newblock
\showISBNx{9781450394215}
\urldef\tempurl%
\url{https://doi.org/10.1145/3544548.3581075}
\showDOI{\tempurl}


\bibitem[Smith et~al\mbox{.}(2013)]%
        {smith2013ability}
\bibfield{author}{\bibinfo{person}{Gary~B Smith}, \bibinfo{person}{David~R Prytherch}, \bibinfo{person}{Paul Meredith}, \bibinfo{person}{Paul~E Schmidt}, {and} \bibinfo{person}{Peter~I Featherstone}.} \bibinfo{year}{2013}\natexlab{}.
\newblock \showarticletitle{The ability of the National Early Warning Score (NEWS) to discriminate patients at risk of early cardiac arrest, unanticipated intensive care unit admission, and death}.
\newblock \bibinfo{journal}{\emph{Resuscitation}} \bibinfo{volume}{84}, \bibinfo{number}{4} (\bibinfo{year}{2013}), \bibinfo{pages}{465--470}.
\newblock


\bibitem[Smith-Renner et~al\mbox{.}(2020)]%
        {10.1145/3313831.3376624}
\bibfield{author}{\bibinfo{person}{Alison Smith-Renner}, \bibinfo{person}{Ron Fan}, \bibinfo{person}{Melissa Birchfield}, \bibinfo{person}{Tongshuang Wu}, \bibinfo{person}{Jordan Boyd-Graber}, \bibinfo{person}{Daniel~S. Weld}, {and} \bibinfo{person}{Leah Findlater}.} \bibinfo{year}{2020}\natexlab{}.
\newblock \showarticletitle{No Explainability without Accountability: An Empirical Study of Explanations and Feedback in Interactive ML}. In \bibinfo{booktitle}{\emph{Proceedings of the 2020 CHI Conference on Human Factors in Computing Systems}} (Honolulu, HI, USA) \emph{(\bibinfo{series}{CHI '20})}. \bibinfo{publisher}{Association for Computing Machinery}, \bibinfo{address}{New York, NY, USA}, \bibinfo{pages}{1–13}.
\newblock
\showISBNx{9781450367080}
\urldef\tempurl%
\url{https://doi.org/10.1145/3313831.3376624}
\showDOI{\tempurl}


\bibitem[Sox et~al\mbox{.}(2013)]%
        {sox2013medical}
\bibfield{author}{\bibinfo{person}{Harold~C Sox}, \bibinfo{person}{Michael~C. Higgins}, {and} \bibinfo{person}{Douglas~K. Owens}.} \bibinfo{year}{2013}\natexlab{}.
\newblock \bibinfo{booktitle}{\emph{Medical Decision Making}}.
\newblock \bibinfo{publisher}{John Wiley \& Sons, Ltd}.
\newblock


\bibitem[Subbe et~al\mbox{.}(2001)]%
        {subbe2001validation}
\bibfield{author}{\bibinfo{person}{Christian~P Subbe}, \bibinfo{person}{Michael Kruger}, \bibinfo{person}{Peter Rutherford}, {and} \bibinfo{person}{L Gemmel}.} \bibinfo{year}{2001}\natexlab{}.
\newblock \showarticletitle{Validation of a modified Early Warning Score in medical admissions}.
\newblock \bibinfo{journal}{\emph{Qjm}} \bibinfo{volume}{94}, \bibinfo{number}{10} (\bibinfo{year}{2001}), \bibinfo{pages}{521--526}.
\newblock


\bibitem[Tanguay-Sela et~al\mbox{.}(2022)]%
        {tanguay2022evaluating}
\bibfield{author}{\bibinfo{person}{Myriam Tanguay-Sela}, \bibinfo{person}{David Benrimoh}, \bibinfo{person}{Christina Popescu}, \bibinfo{person}{Tamara Perez}, \bibinfo{person}{Colleen Rollins}, \bibinfo{person}{Emily Snook}, \bibinfo{person}{Eryn Lundrigan}, \bibinfo{person}{Caitrin Armstrong}, \bibinfo{person}{Kelly Perlman}, \bibinfo{person}{Robert Fratila}, {et~al\mbox{.}}} \bibinfo{year}{2022}\natexlab{}.
\newblock \showarticletitle{Evaluating the perceived utility of an artificial intelligence-powered clinical decision support system for depression treatment using a simulation center}.
\newblock \bibinfo{journal}{\emph{Psychiatry Research}}  \bibinfo{volume}{308} (\bibinfo{year}{2022}), \bibinfo{pages}{114336}.
\newblock


\bibitem[Thomas(2006)]%
        {thomas2006}
\bibfield{author}{\bibinfo{person}{David~R Thomas}.} \bibinfo{year}{2006}\natexlab{}.
\newblock \showarticletitle{A general inductive approach for analyzing qualitative evaluation data}.
\newblock \bibinfo{journal}{\emph{American journal of evaluation}} \bibinfo{volume}{27}, \bibinfo{number}{2} (\bibinfo{year}{2006}), \bibinfo{pages}{237--246}.
\newblock


\bibitem[Ting et~al\mbox{.}(2018)]%
        {ting2018ai}
\bibfield{author}{\bibinfo{person}{Daniel~SW Ting}, \bibinfo{person}{Yong Liu}, \bibinfo{person}{Philippe Burlina}, \bibinfo{person}{Xinxing Xu}, \bibinfo{person}{Neil~M Bressler}, {and} \bibinfo{person}{Tien~Y Wong}.} \bibinfo{year}{2018}\natexlab{}.
\newblock \showarticletitle{AI for medical imaging goes deep}.
\newblock \bibinfo{journal}{\emph{Nature medicine}} \bibinfo{volume}{24}, \bibinfo{number}{5} (\bibinfo{year}{2018}), \bibinfo{pages}{539--540}.
\newblock


\bibitem[Topol(2019)]%
        {topol2019high}
\bibfield{author}{\bibinfo{person}{Eric~J Topol}.} \bibinfo{year}{2019}\natexlab{}.
\newblock \showarticletitle{High-performance medicine: the convergence of human and artificial intelligence}.
\newblock \bibinfo{journal}{\emph{Nature medicine}} \bibinfo{volume}{25}, \bibinfo{number}{1} (\bibinfo{year}{2019}), \bibinfo{pages}{44--56}.
\newblock


\bibitem[Usman et~al\mbox{.}(2019)]%
        {usman2019comparison}
\bibfield{author}{\bibinfo{person}{Omar~A Usman}, \bibinfo{person}{Asad~A Usman}, {and} \bibinfo{person}{Michael~A Ward}.} \bibinfo{year}{2019}\natexlab{}.
\newblock \showarticletitle{Comparison of SIRS, qSOFA, and NEWS for the early identification of sepsis in the Emergency Department}.
\newblock \bibinfo{journal}{\emph{The American journal of emergency medicine}} \bibinfo{volume}{37}, \bibinfo{number}{8} (\bibinfo{year}{2019}), \bibinfo{pages}{1490--1497}.
\newblock


\bibitem[Wang et~al\mbox{.}(2021)]%
        {wang2021brilliant}
\bibfield{author}{\bibinfo{person}{Dakuo Wang}, \bibinfo{person}{Liuping Wang}, \bibinfo{person}{Zhan Zhang}, \bibinfo{person}{Ding Wang}, \bibinfo{person}{Haiyi Zhu}, \bibinfo{person}{Yvonne Gao}, \bibinfo{person}{Xiangmin Fan}, {and} \bibinfo{person}{Feng Tian}.} \bibinfo{year}{2021}\natexlab{}.
\newblock \showarticletitle{“Brilliant AI Doctor” in Rural Clinics: Challenges in AI-Powered Clinical Decision Support System Deployment}. In \bibinfo{booktitle}{\emph{Proceedings of the 2021 CHI Conference on Human Factors in Computing Systems}} (Yokohama, Japan) \emph{(\bibinfo{series}{CHI '21})}. \bibinfo{publisher}{Association for Computing Machinery}, \bibinfo{address}{New York, NY, USA}, Article \bibinfo{articleno}{697}, \bibinfo{numpages}{18}~pages.
\newblock
\showISBNx{9781450380966}
\urldef\tempurl%
\url{https://doi.org/10.1145/3411764.3445432}
\showDOI{\tempurl}


\bibitem[Wang and Yin(2021)]%
        {wang2021explanations}
\bibfield{author}{\bibinfo{person}{Xinru Wang} {and} \bibinfo{person}{Ming Yin}.} \bibinfo{year}{2021}\natexlab{}.
\newblock \showarticletitle{Are Explanations Helpful? A Comparative Study of the Effects of Explanations in AI-Assisted Decision-Making}. In \bibinfo{booktitle}{\emph{26th International Conference on Intelligent User Interfaces}} (College Station, TX, USA) \emph{(\bibinfo{series}{IUI '21})}. \bibinfo{publisher}{Association for Computing Machinery}, \bibinfo{address}{New York, NY, USA}, \bibinfo{pages}{318–328}.
\newblock
\showISBNx{9781450380171}
\urldef\tempurl%
\url{https://doi.org/10.1145/3397481.3450650}
\showDOI{\tempurl}


\bibitem[Wong et~al\mbox{.}(2021)]%
        {wong2021external}
\bibfield{author}{\bibinfo{person}{Andrew Wong}, \bibinfo{person}{Erkin Otles}, \bibinfo{person}{John~P Donnelly}, \bibinfo{person}{Andrew Krumm}, \bibinfo{person}{Jeffrey McCullough}, \bibinfo{person}{Olivia DeTroyer-Cooley}, \bibinfo{person}{Justin Pestrue}, \bibinfo{person}{Marie Phillips}, \bibinfo{person}{Judy Konye}, \bibinfo{person}{Carleen Penoza}, {et~al\mbox{.}}} \bibinfo{year}{2021}\natexlab{}.
\newblock \showarticletitle{External validation of a widely implemented proprietary sepsis prediction model in hospitalized patients}.
\newblock \bibinfo{journal}{\emph{JAMA Internal Medicine}} \bibinfo{volume}{181}, \bibinfo{number}{8} (\bibinfo{year}{2021}), \bibinfo{pages}{1065--1070}.
\newblock


\bibitem[{World Health Organization}(2020)]%
        {world2020global}
\bibfield{author}{\bibinfo{person}{{World Health Organization}}.} \bibinfo{year}{2020}\natexlab{}.
\newblock \showarticletitle{Global report on the epidemiology and burden of sepsis: current evidence, identifying gaps and future directions}.
\newblock  (\bibinfo{year}{2020}).
\newblock


\bibitem[Xiong et~al\mbox{.}(2022)]%
        {xiong2022challenges}
\bibfield{author}{\bibinfo{person}{Wei Xiong}, \bibinfo{person}{Hongmiao Fan}, \bibinfo{person}{Liang Ma}, {and} \bibinfo{person}{Chen Wang}.} \bibinfo{year}{2022}\natexlab{}.
\newblock \showarticletitle{Challenges of human—machine collaboration in risky decision-making}.
\newblock \bibinfo{journal}{\emph{Frontiers of Engineering Management}} \bibinfo{volume}{9}, \bibinfo{number}{1} (\bibinfo{year}{2022}), \bibinfo{pages}{89--103}.
\newblock


\bibitem[Xu et~al\mbox{.}(2019)]%
        {xu2019leveraging}
\bibfield{author}{\bibinfo{person}{Xuhai Xu}, \bibinfo{person}{Prerna Chikersal}, \bibinfo{person}{Afsaneh Doryab}, \bibinfo{person}{Daniella~K. Villalba}, \bibinfo{person}{Janine~M. Dutcher}, \bibinfo{person}{Michael~J. Tumminia}, \bibinfo{person}{Tim Althoff}, \bibinfo{person}{Sheldon Cohen}, \bibinfo{person}{Kasey~G. Creswell}, \bibinfo{person}{J.~David Creswell}, \bibinfo{person}{Jennifer Mankoff}, {and} \bibinfo{person}{Anind~K. Dey}.} \bibinfo{year}{2019}\natexlab{}.
\newblock \showarticletitle{Leveraging Routine Behavior and Contextually-Filtered Features for Depression Detection among College Students}.
\newblock \bibinfo{journal}{\emph{Proc. ACM Interact. Mob. Wearable Ubiquitous Technol.}} \bibinfo{volume}{3}, \bibinfo{number}{3}, Article \bibinfo{articleno}{116} (\bibinfo{date}{sep} \bibinfo{year}{2019}), \bibinfo{numpages}{33}~pages.
\newblock
\urldef\tempurl%
\url{https://doi.org/10.1145/3351274}
\showDOI{\tempurl}


\bibitem[Xu et~al\mbox{.}(2023)]%
        {xu2023globem}
\bibfield{author}{\bibinfo{person}{Xuhai Xu}, \bibinfo{person}{Xin Liu}, \bibinfo{person}{Han Zhang}, \bibinfo{person}{Weichen Wang}, \bibinfo{person}{Subigya Nepal}, \bibinfo{person}{Yasaman Sefidgar}, \bibinfo{person}{Woosuk Seo}, \bibinfo{person}{Kevin~S. Kuehn}, \bibinfo{person}{Jeremy~F. Huckins}, \bibinfo{person}{Margaret~E. Morris}, \bibinfo{person}{Paula~S. Nurius}, \bibinfo{person}{Eve~A. Riskin}, \bibinfo{person}{Shwetak Patel}, \bibinfo{person}{Tim Althoff}, \bibinfo{person}{Andrew Campbell}, \bibinfo{person}{Anind~K. Dey}, {and} \bibinfo{person}{Jennifer Mankoff}.} \bibinfo{year}{2023}\natexlab{}.
\newblock \showarticletitle{GLOBEM: Cross-Dataset Generalization of Longitudinal Human Behavior Modeling}.
\newblock \bibinfo{journal}{\emph{Proc. ACM Interact. Mob. Wearable Ubiquitous Technol.}} \bibinfo{volume}{6}, \bibinfo{number}{4}, Article \bibinfo{articleno}{190} (\bibinfo{date}{jan} \bibinfo{year}{2023}), \bibinfo{numpages}{34}~pages.
\newblock
\urldef\tempurl%
\url{https://doi.org/10.1145/3569485}
\showDOI{\tempurl}


\bibitem[Xu et~al\mbox{.}(2022)]%
        {xu2022globem}
\bibfield{author}{\bibinfo{person}{Xuhai Xu}, \bibinfo{person}{Han Zhang}, \bibinfo{person}{Yasaman Sefidgar}, \bibinfo{person}{Yiyi Ren}, \bibinfo{person}{Xin Liu}, \bibinfo{person}{Woosuk Seo}, \bibinfo{person}{Jennifer Brown}, \bibinfo{person}{Kevin Kuehn}, \bibinfo{person}{Mike Merrill}, \bibinfo{person}{Paula Nurius}, \bibinfo{person}{Shwetak Patel}, \bibinfo{person}{Tim Althoff}, \bibinfo{person}{Margaret Morris}, \bibinfo{person}{Eve Riskin}, \bibinfo{person}{Jennifer Mankoff}, {and} \bibinfo{person}{Anind Dey}.} \bibinfo{year}{2022}\natexlab{}.
\newblock \showarticletitle{GLOBEM Dataset: Multi-Year Datasets for Longitudinal Human Behavior Modeling Generalization}. In \bibinfo{booktitle}{\emph{Advances in Neural Information Processing Systems}}, \bibfield{editor}{\bibinfo{person}{S.~Koyejo}, \bibinfo{person}{S.~Mohamed}, \bibinfo{person}{A.~Agarwal}, \bibinfo{person}{D.~Belgrave}, \bibinfo{person}{K.~Cho}, {and} \bibinfo{person}{A.~Oh}} (Eds.), Vol.~\bibinfo{volume}{35}. \bibinfo{publisher}{Curran Associates, Inc.}, \bibinfo{pages}{24655--24692}.
\newblock


\bibitem[Yang et~al\mbox{.}(2023)]%
        {yang2023harnessing}
\bibfield{author}{\bibinfo{person}{Qian Yang}, \bibinfo{person}{Yuexing Hao}, \bibinfo{person}{Kexin Quan}, \bibinfo{person}{Stephen Yang}, \bibinfo{person}{Yiran Zhao}, \bibinfo{person}{Volodymyr Kuleshov}, {and} \bibinfo{person}{Fei Wang}.} \bibinfo{year}{2023}\natexlab{}.
\newblock \showarticletitle{Harnessing Biomedical Literature to Calibrate Clinicians’ Trust in AI Decision Support Systems}. In \bibinfo{booktitle}{\emph{Proceedings of the 2023 CHI Conference on Human Factors in Computing Systems}} (Hamburg, Germany) \emph{(\bibinfo{series}{CHI '23})}. \bibinfo{publisher}{Association for Computing Machinery}, \bibinfo{address}{New York, NY, USA}, Article \bibinfo{articleno}{14}, \bibinfo{numpages}{14}~pages.
\newblock
\showISBNx{9781450394215}
\urldef\tempurl%
\url{https://doi.org/10.1145/3544548.3581393}
\showDOI{\tempurl}


\bibitem[Yang et~al\mbox{.}(2019)]%
        {yang2019unremarkable}
\bibfield{author}{\bibinfo{person}{Qian Yang}, \bibinfo{person}{Aaron Steinfeld}, {and} \bibinfo{person}{John Zimmerman}.} \bibinfo{year}{2019}\natexlab{}.
\newblock \showarticletitle{Unremarkable AI: Fitting intelligent decision support into critical, clinical decision-making processes}. In \bibinfo{booktitle}{\emph{Proceedings of the 2019 CHI conference on human factors in computing systems}}. \bibinfo{pages}{1--11}.
\newblock


\bibitem[Yang et~al\mbox{.}(2016)]%
        {yang2016investigating}
\bibfield{author}{\bibinfo{person}{Qian Yang}, \bibinfo{person}{John Zimmerman}, \bibinfo{person}{Aaron Steinfeld}, \bibinfo{person}{Lisa Carey}, {and} \bibinfo{person}{James~F Antaki}.} \bibinfo{year}{2016}\natexlab{}.
\newblock \showarticletitle{Investigating the heart pump implant decision process: opportunities for decision support tools to help}. In \bibinfo{booktitle}{\emph{Proceedings of the 2016 CHI Conference on Human Factors in Computing Systems}}. \bibinfo{pages}{4477--4488}.
\newblock


\bibitem[Yin et~al\mbox{.}(2022)]%
        {yin2022deconfounding}
\bibfield{author}{\bibinfo{person}{Changchang Yin}, \bibinfo{person}{Ruoqi Liu}, \bibinfo{person}{Jeffrey Caterino}, {and} \bibinfo{person}{Ping Zhang}.} \bibinfo{year}{2022}\natexlab{}.
\newblock \showarticletitle{Deconfounding Actor-Critic Network with Policy Adaptation for Dynamic Treatment Regimes}. In \bibinfo{booktitle}{\emph{Proceedings of the 28th ACM SIGKDD Conference on Knowledge Discovery and Data Mining}} (Washington DC, USA) \emph{(\bibinfo{series}{KDD '22})}. \bibinfo{publisher}{Association for Computing Machinery}, \bibinfo{address}{New York, NY, USA}, \bibinfo{pages}{2316–2326}.
\newblock
\showISBNx{9781450393850}
\urldef\tempurl%
\url{https://doi.org/10.1145/3534678.3539413}
\showDOI{\tempurl}


\bibitem[Yin et~al\mbox{.}(2020)]%
        {yin2020identifying}
\bibfield{author}{\bibinfo{person}{Changchang Yin}, \bibinfo{person}{Ruoqi Liu}, \bibinfo{person}{Dongdong Zhang}, {and} \bibinfo{person}{Ping Zhang}.} \bibinfo{year}{2020}\natexlab{}.
\newblock \showarticletitle{Identifying Sepsis Subphenotypes via Time-Aware Multi-Modal Auto-Encoder}. In \bibinfo{booktitle}{\emph{Proceedings of the 26th ACM SIGKDD International Conference on Knowledge Discovery \& Data Mining}} (Virtual Event, CA, USA) \emph{(\bibinfo{series}{KDD '20})}. \bibinfo{publisher}{Association for Computing Machinery}, \bibinfo{address}{New York, NY, USA}, \bibinfo{pages}{862–872}.
\newblock
\showISBNx{9781450379984}
\urldef\tempurl%
\url{https://doi.org/10.1145/3394486.3403129}
\showDOI{\tempurl}


\bibitem[Yin et~al\mbox{.}(2019)]%
        {yin2019domain}
\bibfield{author}{\bibinfo{person}{Changchang Yin}, \bibinfo{person}{Rongjian Zhao}, \bibinfo{person}{Buyue Qian}, \bibinfo{person}{Xin Lv}, {and} \bibinfo{person}{Ping Zhang}.} \bibinfo{year}{2019}\natexlab{}.
\newblock \showarticletitle{Domain knowledge guided deep learning with electronic health records}. In \bibinfo{booktitle}{\emph{2019 IEEE International Conference on Data Mining (ICDM)}}. IEEE, \bibinfo{pages}{738--747}.
\newblock


\bibitem[Yu et~al\mbox{.}(2018)]%
        {yu2018artificial}
\bibfield{author}{\bibinfo{person}{Kun-Hsing Yu}, \bibinfo{person}{Andrew~L Beam}, {and} \bibinfo{person}{Isaac~S Kohane}.} \bibinfo{year}{2018}\natexlab{}.
\newblock \showarticletitle{Artificial intelligence in healthcare}.
\newblock \bibinfo{journal}{\emph{Nature biomedical engineering}} \bibinfo{volume}{2}, \bibinfo{number}{10} (\bibinfo{year}{2018}), \bibinfo{pages}{719--731}.
\newblock


\bibitem[Yuan et~al\mbox{.}(2020)]%
        {yuan2020development}
\bibfield{author}{\bibinfo{person}{Kuo-Ching Yuan}, \bibinfo{person}{Lung-Wen Tsai}, \bibinfo{person}{Ko-Han Lee}, \bibinfo{person}{Yi-Wei Cheng}, \bibinfo{person}{Shou-Chieh Hsu}, \bibinfo{person}{Yu-Sheng Lo}, {and} \bibinfo{person}{Ray-Jade Chen}.} \bibinfo{year}{2020}\natexlab{}.
\newblock \showarticletitle{The development an artificial intelligence algorithm for early sepsis diagnosis in the intensive care unit}.
\newblock \bibinfo{journal}{\emph{International Journal of Medical Informatics}}  \bibinfo{volume}{141} (\bibinfo{year}{2020}), \bibinfo{pages}{104176}.
\newblock
\showISSN{1386-5056}
\urldef\tempurl%
\url{https://doi.org/10.1016/j.ijmedinf.2020.104176}
\showDOI{\tempurl}


\bibitem[Zaj\k{a}c et~al\mbox{.}(2023)]%
        {10.1145/3582430}
\bibfield{author}{\bibinfo{person}{Hubert~D. Zaj\k{a}c}, \bibinfo{person}{Dana Li}, \bibinfo{person}{Xiang Dai}, \bibinfo{person}{Jonathan~F. Carlsen}, \bibinfo{person}{Finn Kensing}, {and} \bibinfo{person}{Tariq~O. Andersen}.} \bibinfo{year}{2023}\natexlab{}.
\newblock \showarticletitle{Clinician-Facing AI in the Wild: Taking Stock of the Sociotechnical Challenges and Opportunities for HCI}.
\newblock \bibinfo{journal}{\emph{ACM Trans. Comput.-Hum. Interact.}} \bibinfo{volume}{30}, \bibinfo{number}{2}, Article \bibinfo{articleno}{33} (\bibinfo{date}{mar} \bibinfo{year}{2023}), \bibinfo{numpages}{39}~pages.
\newblock
\showISSN{1073-0516}
\urldef\tempurl%
\url{https://doi.org/10.1145/3582430}
\showDOI{\tempurl}


\bibitem[Zambon et~al\mbox{.}(2008)]%
        {zambon2008implementation}
\bibfield{author}{\bibinfo{person}{Massimo Zambon}, \bibinfo{person}{Marcello Ceola}, \bibinfo{person}{Roberto~Almeida de Castro}, \bibinfo{person}{Antonino Gullo}, {and} \bibinfo{person}{Jean-Louis Vincent}.} \bibinfo{year}{2008}\natexlab{}.
\newblock \showarticletitle{Implementation of the Surviving Sepsis Campaign guidelines for severe sepsis and septic shock: We could go faster}.
\newblock \bibinfo{journal}{\emph{Journal of Critical Care}} \bibinfo{volume}{23}, \bibinfo{number}{4} (\bibinfo{year}{2008}), \bibinfo{pages}{455--460}.
\newblock
\showISSN{0883-9441}
\urldef\tempurl%
\url{https://doi.org/10.1016/j.jcrc.2007.08.003}
\showDOI{\tempurl}


\bibitem[Zhang et~al\mbox{.}(2021)]%
        {zhang2021interpretable}
\bibfield{author}{\bibinfo{person}{Dongdong Zhang}, \bibinfo{person}{Changchang Yin}, \bibinfo{person}{Katherine~M. Hunold}, \bibinfo{person}{Xiaoqian Jiang}, \bibinfo{person}{Jeffrey~M. Caterino}, {and} \bibinfo{person}{Ping Zhang}.} \bibinfo{year}{2021}\natexlab{}.
\newblock \showarticletitle{An interpretable deep-learning model for early prediction of sepsis in the emergency department}.
\newblock \bibinfo{journal}{\emph{Patterns}} \bibinfo{volume}{2}, \bibinfo{number}{2} (\bibinfo{year}{2021}), \bibinfo{pages}{100196}.
\newblock
\showISSN{2666-3899}
\urldef\tempurl%
\url{https://doi.org/10.1016/j.patter.2020.100196}
\showDOI{\tempurl}


\bibitem[Zhang et~al\mbox{.}(2020)]%
        {zhang2020effect}
\bibfield{author}{\bibinfo{person}{Yunfeng Zhang}, \bibinfo{person}{Q.~Vera Liao}, {and} \bibinfo{person}{Rachel K.~E. Bellamy}.} \bibinfo{year}{2020}\natexlab{}.
\newblock \showarticletitle{Effect of Confidence and Explanation on Accuracy and Trust Calibration in AI-Assisted Decision Making}. In \bibinfo{booktitle}{\emph{Proceedings of the 2020 Conference on Fairness, Accountability, and Transparency}} (Barcelona, Spain) \emph{(\bibinfo{series}{FAT* '20})}. \bibinfo{publisher}{Association for Computing Machinery}, \bibinfo{address}{New York, NY, USA}, \bibinfo{pages}{295–305}.
\newblock
\showISBNx{9781450369367}
\urldef\tempurl%
\url{https://doi.org/10.1145/3351095.3372852}
\showDOI{\tempurl}


\bibitem[Zhang et~al\mbox{.}(2017)]%
        {zhang2017lstm}
\bibfield{author}{\bibinfo{person}{Yuan Zhang}, \bibinfo{person}{Chen Lin}, \bibinfo{person}{Min Chi}, \bibinfo{person}{Julie Ivy}, \bibinfo{person}{Muge Capan}, {and} \bibinfo{person}{Jeanne~M. Huddleston}.} \bibinfo{year}{2017}\natexlab{}.
\newblock \showarticletitle{LSTM for septic shock: Adding unreliable labels to reliable predictions}. In \bibinfo{booktitle}{\emph{2017 IEEE International Conference on Big Data (Big Data)}}. \bibinfo{pages}{1233--1242}.
\newblock
\urldef\tempurl%
\url{https://doi.org/10.1109/BigData.2017.8258049}
\showDOI{\tempurl}


\bibitem[Zhu et~al\mbox{.}(2016)]%
        {zhu2016measuring}
\bibfield{author}{\bibinfo{person}{Zihao Zhu}, \bibinfo{person}{Changchang Yin}, \bibinfo{person}{Buyue Qian}, \bibinfo{person}{Yu Cheng}, \bibinfo{person}{Jishang Wei}, {and} \bibinfo{person}{Fei Wang}.} \bibinfo{year}{2016}\natexlab{}.
\newblock \showarticletitle{Measuring Patient Similarities via a Deep Architecture with Medical Concept Embedding}. In \bibinfo{booktitle}{\emph{2016 IEEE 16th International Conference on Data Mining (ICDM)}}. \bibinfo{pages}{749--758}.
\newblock
\urldef\tempurl%
\url{https://doi.org/10.1109/ICDM.2016.0086}
\showDOI{\tempurl}


\end{thebibliography}
\clearpage
\appendix

\section{Formative Study Interview Script}\label{appendix:interview_script}
\begin{itemize}
    \item \textbf{Question 1 - Background:} Can you tell us a bit more about your work, such as your job title, years of practice, main working context, etc.?
    \item \textbf{Question 2 - Experience of Sepsis Diagnosis:} Could you recall a recent sepsis encounter that stands out to you? When/where it happened? (Please do not mention any patient personally identifiable information)
    \item \textbf{Question 3 - Experience of Existing Epic Sepsis Module:} In the previous sepsis encounter, how did you use the sepsis diagnosis module in Epic EHR? Useful or not at all?
    \item \textbf{Question 4 - Attitudes towards AI Diagnosis:} How do you like the sepsis risks score predicted by AI model? [Do you want AI to assisted you?] [What information you wish AI could have provided you?]
    \item \textbf{Question 5 - Closing Question:} Is there anything else that you would like to share with us or any questions you have for us?

\end{itemize}

\section{LSTM Model Performance}\label{app:LSTMperformace}

To demonstrate the performance of our chosen LSTM model, we have excerpted its performance at Figure \ref{tab:classification_results} on two different cases from \cite{zhang2021interpretable}.
\paragraph*{Case 1} In this case, patients are sampled from septic patients, and the goal is to see if a model can tell if a patient is likely to have high sepsis risk a few hours before the onset. For each patient, the patient records is split into 2 segments at the middle point, segment close to sepsis onset ( = 4 hours) is labeled as 1, another segment ( $>$ 4 hours before sepsis onset) is labeled 0. We randomly pick either the former or latter segment to build the Case1 cohort. The introduction of case 1 is to measure the model in terms of time-sensitive prediction to ensure models are indeed clinically useful and relieve "warning fatigue" as alarm burden. Given patient records either from $T_{admission}$ to $T_{middle}$ or from $T_{middle}$ to $T_{onset}-4$, our model is required to distinguish this 2 kinds of records.
\paragraph*{Case 2} In this case, case and control segments are from different patients who have sepsis onset in the next 4 hours, as well as those who do not have sepsis. Given patient records from $T_{admission}$ to $T_{onset}-4$, we are going to predict whether sepsis occurs in the following 4 hours.

\begin{table}[t]
    \centering
    \caption{AUC accuracy scores of sepsis prediction tasks on DII-challenge dataset from \cite{zhang2021interpretable}. }
    \begin{tabular}{lccc}
        \toprule
        \textbf{Method} & \textbf{Case1} & \textbf{Case2} & \textbf{Average} \\ \midrule
        MEWS & 0.54 & 0.72 & 0.63 \\
        NEWS & 0.52 & 0.72 & 0.62 \\
        SIRS & 0.56 & 0.69 & 0.62 \\
        qSOFA & 0.53 & 0.65 & 0.59 \\ \midrule
        Logistic regression & 0.89 & 0.79 & 0.84 \\
        Random forest & 0.90 & 0.81 & 0.85 \\
        Gradient boosting trees & 0.91 & 0.81 & 0.86 \\ \midrule
        GRU & 0.88 & 0.80 & 0.84 \\ 
        RETAIN \cite{choi2016retain} & 0.90 & 0.80 & 0.85 \\
        Dipole \cite{ma2017dipole} & 0.90 & 0.81 & 0.86 \\ \midrule
        LSTM \cite{zhang2021interpretable} & \textbf{0.94} & \textbf{0.84} & \textbf{0.89} \\
        \bottomrule
    \end{tabular}
    \label{tab:classification_results}
\end{table}

\section{Model Parameters}\label{app:modelpara}

\paragraph{Sepsis onset risk prediction model. }Following~\cite{zhang2021interpretable},  we used PyTorch \cite{paszke2019pytorch} to implement the LSTM-based sepsis risk prediction model, and the number of timesteps for LSTM was set to 100. The sizes of variable embedding vectors and hidden state vectors were set as 256. The number of layers in LSTM was set as 2. 
For evaluation, 80\% of the data are used for training, and 10\% for validation, 10\% for testing. The results are displayed in \autoref{tab:classification_results}.

\paragraph{Recommendation algorithm.} Two thresholds $th_s$ and $th_e$ ($1>th_s, th_e>0$) are used to decide whether new laboratory values should be requested in \autoref{uncertainty_estimation}. The prediction threshold $th_s$ was set as 0.5 to predict whether the patients will have sepsis onset in next 4 hours. The uncertainty threshold $th_e$ was set as 0.25 to determine whether to collect more clinical variables (e.g., to request additional lab tests when Shannon entropy \cite{shannon1949mathematical} uncertainty is bigger than 0.25).
The results are displayed in \autoref{tab:active_sensing_results}.

\section{User Study Interview Script}\label{appendix:userstudy}
\begin{itemize}
    \item \textbf{Question 1 - Attitude to AI Technology in Clinical Scenario:} Do you want any AI in your work?
    \item \textbf{Question 2 - Feedback on Our Initial Design (after showing the screenshot/demo of the UI): }How do you like the design? 
    \item \textbf{Question 3 - Willingness to use:} Would you use such an AI system everyday? 
    \item \textbf{Question 4 - Trust:} Would you trust it? 
    \item \textbf{Question 5 - Improvement}: What improvement do you need?
\end{itemize}

\end{document}